\documentclass[twocolumn,aps,showpacs,prb,tightenlines,amsmath,amssymb]{revtex4}
\usepackage{graphicx}
\usepackage{amssymb}
\usepackage{slashed}
\usepackage{dcolumn}
\usepackage{amsmath}
\usepackage{bm}
\usepackage{colordvi}
\usepackage{mathrsfs}
\makeatletter

\newcommand{\Rmnum}[1]{\expandafter\@slowromancap\romannumeral #1@}
\makeatother

\begin{document}

\title{Influence of scattering on optical response of superconductivity}

\author{F. Yang}
\affiliation{Hefei National Laboratory for Physical Sciences at
Microscale, Department of Physics, and CAS Key Laboratory of Strongly-Coupled
Quantum Matter Physics, University of Science and Technology of China, Hefei,
Anhui, 230026, China}

\author{M. W. Wu}
\thanks{Author to whom correspondence should be addressed}
\email{mwwu@ustc.edu.cn.}

\affiliation{Hefei National Laboratory for Physical Sciences at
Microscale, Department of Physics, and CAS Key Laboratory of Strongly-Coupled
Quantum Matter Physics, University of Science and Technology of China, Hefei,
Anhui, 230026, China}

\date{\today}

\begin{abstract} 
By using the gauge-invariant kinetic equation [Phys. Rev. B {\bf 98}, 094507 (2018); Phys. Rev. B {\bf 100}, 104513 (2019)], we analytically investigate the influence of the scattering on
the optical properties of superconductors in the normal-skin-effect region. 
Both linear and second-order responses are studied under a multi-cycle terahertz pulse. In the linear regime, we reveal that the optical absorption
$\sigma_{1s}(\omega)$, induced by the scattering, exhibits a crossover
point at $\omega=2|\Delta|$. Particularly, 
it is further shown that when $\omega<2|\Delta|$, $\sigma_{1s}(\omega)$ from the scattering always exhibits a finite
value even at low temperature, in contrast to the vanishing $\sigma_{1s}(\omega)$ in the anomalous-skin-effect region as
the Mattis-Bardeen theory [Phys. Rev. {\bf 111}, 412 (1958)] revealed.
In the second-order regime, responses of the Higgs mode during and after the optical pulse
are studied. During the pulse, we show that the scattering 
causes a phase shift in the second-order response of the Higgs mode. 
Particularly, this phase shift exhibits a significant $\pi$-jump at $\omega=|\Delta|$, which provides a very clear feature
for the experimental detection. After the pulse, by studying the damping of the Higgs-mode excitation, we reveal a
relaxation mechanism from the elastic scattering, which shows a monotonic enhancement with the increase of 
the impurity density. 

\end{abstract}


\maketitle

\section{Introduction}

In the past few decades, 
the optical properties of the superconducting states have attracted
much attention in both linear and nonlinear regimes.
The linear response is focused on the behavior of the
optical conductivity \cite{L1,L2,L3,L4,L5,L6,L7,L8,NSL0,NSL1,NSL2,NSL3,NSL4,NSL5,NSL6,NSL8,NSL9,NSL10,NSL11,NSL12}, which was first discussed by 
Mattis and Bardeen (MB)
within the framework of the Kubo current-current correlation approach in the anomalous-skin-effect region \cite{MB,MBo}.
In this region,
the excited current at one space point, depends not only on the
electric field at that point but also on the ones nearby. 
This non-local effect dominates in systems with a small skin depth $\delta$ in comparison with the mean free path $l$,
as usually the
case in thin-film superconductors or clean type-I superconductors, whereas the scattering effect 
in this circumstance
is marginal. The MB theory suggests that the optical absorption at zero temperature is realized by breaking
the Cooper pairs into quasiparticles when the optical frequency $\omega$ is larger than twice the superconducting gap
amplitude $|\Delta|$ \cite{MB}. Thus, the real part of the optical conductivity $\sigma_{1s}(\omega)$ vanishes at {\small
  $T=0~$}K when {\small $\omega<2|\Delta|$} but becomes finite
above $2|\Delta|$, leading to a crossover point at $2|\Delta|$. At finite temperature, an additional quasiparticle contribution appears below $2|\Delta|$. 
This theory so far has successfully described the observed data in the anomalous-skin-effect region, as experiments in
In \cite{L1}, Pb \cite{L2,L6,L7}, Al \cite{L8},
thin-film Nb \cite{L5} and NbN \cite{L3,L4,NSL0}
superconductors demonstrated.
 
The counterpart of the anomalous-skin-effect region is known as 
the normal-skin-effect one \cite{NSL0,NS} ({\small $l<\delta$})
where the dirty type-II superconductors lie in and 
the scattering effect becomes important.
The optical absorption in the normal-skin-effect region, 
as experiments in 
dirty Nb \cite{NSL0,NSL2},  
MgB$_2$ \cite{NSL3,NSL4}, NbTiN \cite{NSL5,NSL6}, NbN \cite{NSL10,NSL8},  MoN \cite{NSL11}
and Al \cite{NSL9,NSL12} superconductors, always exhibits a
finite $\sigma_{1s}(\omega)$ even at low temperature
for {\small $\omega<2|\Delta|$},
in contrast to the vanishing $\sigma_{1s}(\omega)$ in the anomalous-skin-effect region.
Moreover, with the decrease of $\omega$ in terahertz (THz) regime from $\omega\gg2|\Delta|$, 
the observed $\sigma_{1s}(\omega)$ first decreases at {\small $\omega>2|\Delta|$} and then shows an
upturn below $2|\Delta|$, leading to a crossover point at $2|\Delta|$. 
Although the experimental observations are very convincing, 
theories in the normal-skin-effect region where the scattering effect dominates, are
still in progress.  The difficulty within the Kubo formalism comes from 
the inevitable calculation of the vertex correction due to the scattering,
which becomes hard to tackle in superconductors \cite{NS,G1}.  
Whereas the Eilenberger equation is 
restricted by the normalization condition \cite{Eilen,Ba7,Ba8,Ba20}, and   
is also hard to handle for calculation of the scattering. 
So far, to fit the experimental data, the MB theory 
derived from the anomalous region
is excessively used \cite{NSL0,NSL2,NSL4,NSL5,NSL6,NSL8,NSL9,NSL10,NSL11,NSL12}. Nevertheless,
such an unphysical fit
underestimates $\sigma_{1s}(\omega)$ below 
$2|\Delta|$ particularly at low temperature where the 
quasiparticle contribution from MB theory is too small to count for finite experimental result \cite{NSL0,NSL2,NSL4,NSL5,NSL6,NSL8,NSL9,NSL10,NSL11,NSL12}.
To 
explain the residual $\sigma_{1s}(\omega)$, several works \cite{NSL8,NSL9,NSL12} considered the influences of the
collective gapful Higgs \cite{Am1,Am2,Am3,Apm,Am8} and
gapless Nambu-Goldstone \cite{Apm,Am8,gi0,AK,Gm1,Gm2,Ba0,pm0,pm1,pm2,pm3,gi1,Ba9,Ba10} (NG) modes which describe the 
amplitude and phase fluctuations of the order parameter respectively. However,
the Higgs mode is charge neutral and does not 
manifest itself in the linear regime \cite{Am1,Apm,Am8} unless
    under the dc supercurrent injection \cite{DCSI}.
The linear response of the NG mode 
does not occur either due to its coupling with the long-range Coulomb interaction \cite{AK,Apm,pm0,gi1,Ba0,Am1,gi0} 
which causes the original gapless energy lifted up to the
plasma frequency
as a result of Anderson-Higgs mechanism \cite{AHM}. 
Therefore, a detailed study capable of clarifying the scattering effect is necessary. 

As for the non-linear regime, it was recently realized that through 
the intense THz pulse, one can excite the oscillation of the superfluid density in the second-order response, which is
attributed to the excitation of the Higgs mode \cite{NL1,NL2,NL3,NL4,NL5,NL6,NL7}. The most convincing evidence comes from the observed
resonance at $2\omega=2|\Delta|$ \cite{NL2,NL3,NL4}, in consistency with the energy spectrum of the Higgs
mode \cite{Am1,Am3,Apm}. After the THz pulse, a fast damping of this
oscillation is observed, and then, a suppressed gap 
is further observed as a consequence of the thermal effect \cite{NL1,NL2,NL3}.
Theory in the literature for these findings is based on 
Bloch \cite{NL3,NL4,NL5,NL6,NL7,B1,B2,B3,B4,B5,B6} or Liouville \cite{Liou1,Liou2,Liou3,Liou4} equation derived in the
Anderson pseudospin picture \cite{As}. 
The vector potential ${\bf A}$ naturally involves in this description
as a second-order term, which pumps up the fluctuation
of the order parameter (pump effect).
Nevertheless, the microscopic scattering is absent in the literature. 
In order to describe the observed damping after the optical pulse,  
the phenomenological relaxation time is further introduced into the Anderson pseudospin picture \cite{NL6,NL7}.  Very recently,
this whole set of approach is challenged. On one hand, 
this approach with no drive effect fails in the linear regime
to give the optical current. On the other hand, 
symmetry analysis from the Anderson pseudospin picture implies that the pump effect 
excites the NG mode rather than the observed Higgs mode \cite{symmetry}. 
Besides these deficiencies, without the 
microscopic origin, the introduced phenomenological relaxation mechanism is not exact and convincing. 

Very recently, by using the equal-time non-equilibrium
$\tau_0$-Green function, the gauge-invariant kinetic equation (GIKE) of superconductivity 
with the microscopic scattering is developed in our previous
papers \cite{GOBE1,GOBE2,GOBE3,GOBE4,GOBE5}.
We have proved that the retained gauge invariance in this theory directly
leads to the charge conservation in the electromagnetic response \cite{GOBE5},
in consistency with Nambu's conclusion that the gauge invariance in superconductors is equivalent to the charge
conservation \cite{gi0}.
In fact, neither the Bloch \cite{NL3,NL4,NL5,NL6,NL7,B1,B2,B3,B4,B5,B6} nor
Liouville \cite{Liou1,Liou2,Liou3,Liou4} equation mentioned above
are gauge invariant under the gauge transformation in superconductors \cite{gi0}. 
In contrast, in the GIKE, thanks to the gauge invariance, both pump and drive effects mentioned above are kept
\cite{GOBE1,GOBE2,GOBE3,GOBE4,GOBE5}. Moreover, both superfluid and normal-fluid dynamics are involved in the GIKE \cite{GOBE4,GOBE5}, 
beyond the previous Boltzmann equation of superconductors with only the quasiparticle physics retained \cite{Ba3,Bol,Ba5}. 

Consequently, the well-known clean-limit results such as the Ginzburg-Landau equation and Meissner
supercurrent in the magnetic response as well as the optical current captured by the two-fluid model can be directly
derived from the GIKE \cite{GOBE4}.
Particularly, we show that 
the normal fluid is present only when the excited superconducting velocity $v_s$ is larger than a threshold \cite{GOBE4}.
Moreover, the linear responses of 
the collective modes from the GIKE also agree with the 
 well-known results in the literature \cite{GOBE5}. 
Whereas the second-order response from the GIKE exhibits interesting physics. On one hand, a finite second-order
response of the Higgs mode, attributed solely to the drive effect rather than the widely
considered pump effect, is revealed \cite{GOBE5},
in contrast to the above theory from Anderson pseudospin picture \cite{NL3,NL4,NL5,NL6,NL7,B1,B2,B3,B4,B5,B6,Liou1,Liou2,Liou3,Liou4}. On
the other hand, a finite second-order response of the NG mode, survived from 
the Anderson-Higgs mechanism, is predicted as a consequence of charge conservation. An experimental
scheme for this response is further proposed \cite{GOBE5}. 
Actually, thanks to the equal-time scheme, the microscopic scattering in superconductors, which is hard to
deal with in the literature as mentioned above, becomes easy to handle within the GIKE approach. Thus, rich physics from
the scattering can be expected. Particularly, at low frequency (i.e., large $v_s$), we have analytically shown that due to the scattering, 
there exists viscous superfluid besides the non-viscous one \cite{GOBE4}. Then, together with the normal fluid, a
three-fluid model is proposed \cite{GOBE4}. 

In this work, by extending the previous scattering terms in Ref.~\onlinecite{GOBE4}
into the THz regime via carefully implementing the Markovian approximation,
we further apply the GIKE to investigate the influence of the scattering on the
optical properties of superconductors in the normal-skin-effect region ($l<\delta$). Both linear and second-order
responses are analytically studied under a multi-cycle THz pulse.
In the linear regime, we show that the optical absorption
$\sigma_{1s}(\omega)$, induced by the
scattering, always exhibits a finite value even at low temperature
when $\omega<2|\Delta|$, in contrast to the vanishing
$\sigma_{1s}(\omega)$ in the anomalous-skin-effect region as MB theory
revealed \cite{MB}.    
Moreover, with the
decrease of the optical frequency from 
$\omega\gg2|\Delta|$, $\sigma_{1s}(\omega)$ first increases and then
drops abruptly around $2|\Delta|$. By further decreasing $\omega$ below $2|\Delta|$, 
an upturn of $\sigma_{1s}(\omega)$ is observed, leading to a crossover point at $2|\Delta|$.
In the second-order regime, responses of the Higgs mode during and after the optical pulse
are revealed.  During the pulse, it is found that the scattering 
causes a phase shift in the optical response of the Higgs mode. 
Particularly, this phase shift exhibits a significant $\pi$-jump at
$\omega=|\Delta|$, which provides a very clear feature for the experimental detection.
After the pulse, the damping
of the Higgs-mode excitation is studied. In this situation, 
we reveal a relaxation mechanism
due to the elastic scattering, which shows a monotonic
enhancement with the increase of the impurity density. 

This paper is organized as follows. We first present the GIKE of superconductivity in
Sec.~II. Then, we perform the analytic analysis of the influence from the scattering on the 
optical properties of superconductors in Sec.~III. We summarize in Sec. IV.

\section{MODEL}
\label{model}

In this section, we first introduce the complete GIKE. 
Then, we present a simplified GIKE to study the optical response of superconductors
in the normal-skin-effect region. The microscopic scattering terms of the non-magnetic impurity scattering are also addressed in this section.  

\subsection{GIKE}

The GIKE of the $s$-wave BCS superconductors, which is developed in our previous papers \cite{GOBE4,GOBE5}, reads:
\begin{eqnarray}
&&\partial_t\rho^c_{\bf
  k}\!+\!i\Big[(\xi_k\!+\!e\phi\!+\!\mu_H+\!\mu_F)\tau_3\!+{\hat
      \Delta({R})},\rho^c_{\bf
  k}\Big]\!\nonumber\\
&&\mbox{}\!+\!i\Big[\frac{e^2A^2}{2m}\tau_3,\rho^c_{\bf k}\Big]
\!-\!i\Big[\frac{1}{8m}\tau_3,{\bm \nabla}^2_{\bf R}\rho^c_{\bf
k}\Big]\!+\frac{1}{2}\!\Big\{\frac{\bf k}{m}\tau_3,{\bm \nabla}_{\bf R}\rho^c_{\bf
    k}\Big\}\nonumber\\
&&\mbox{}+\!\frac{1}{2}\left\{e{\bf E}\tau_3\!-\!({\bm \nabla}_{\bf R}\!-\!2ie{\bf A}\tau_3){\hat
  \Delta}(R),\partial_{\bf k}\rho^c_{\bf
k}\right\}\nonumber\\
&&\mbox{}\!-\!\frac{i}{8}\Big[({\bm \nabla_{\bf R}}\!-\!2ie{\bf A}\tau_3)({\bm
    \nabla_{\bf R}}\!-\!2ie{\bf A}\tau_3){\hat
  \Delta}(R),\partial_{\bf k}\partial_{\bf k}\rho^c_{\bf
k}\Big]\nonumber\\
&&\mbox{}\!-\!e\Big[\frac{2{\bf
    A}\cdot{\bm \nabla}_{\bf R}\!+\!{\bm
      \nabla}_{\bf R}\cdot{\bf
    A}}{4m}\tau_3,\tau_3\rho^c_{\bf
    k}\Big]\!=\!\partial_t\rho^c_{\bf k}\Big|_{\rm
sc}.
\label{KE}
\end{eqnarray}
Here, $[~,~]$ and $\{~,~\}$ represent the commutator and anti-commutator, 
respectively; $\xi_{k}=\frac{k^2}{2m}-\mu$ with $m$ and $\mu$  
being the effective mass and chemical potential; {\small ${\hat
    \Delta(R)}=\Delta(R)\tau_++\Delta^*(R)\tau_-$};   
$R=(t,{\bf R})$ stands for the center-of-mass coordinate;
$\tau_i$ are the Pauli matrices in the particle-hole space;
${\phi}$ and ${\bf A}$ denote the scalar and vector potentials, respectively;
$\rho^c_{\bf k}$
is the density matrix in the Nambu space; on the right-hand side of  
Eq.~(\ref{KE}), the scattering term $\partial_t\rho^c_{\bf k}\Big|_{\rm sc}$
is added for the completeness, whose explicit expression is shown in Sec.~\ref{sesc}. 

The superconducting order parameter $\Delta$, Fock field $\mu_F$
and Hartree field $\mu_H$ in Eq.~(\ref{KE}) are written as 
\begin{eqnarray}
\Delta(R)&=&-g{\sum_{\bf k}}'{\rm Tr}[\rho^c_{\bf k}\tau_-],\label{dp}\\
\Delta^*(R)&=&-g{\sum_{\bf k}}'{\rm Tr}[\rho^c_{\bf k}\tau_+],\label{dm}\\
\mu_F(R)&=&g\delta{n(R)}/2,\label{muFn}\\
\mu_H(R)&=&\sum_{\bf R'}V_{{\bf R}-{\bf
    R'}}{\delta}n({\bf R}'),\label{muH}
\end{eqnarray}
where $\delta{n(R)}$ represents the density fluctuation;
$V_{\bf R-R'}$ denotes the Coulomb potential whose Fourier component $V_{\bf 
  q}=e^2/(q^2\epsilon_0)$; 
 $g$ stands for the effective electron-electron attractive potential in the BCS
theory \cite{BCS}. $\sum'_{\bf k}$ here and hereafter represents the summation restricted in the spherical shell
($|\xi_k|<\omega_D$) with $\omega_D$ being the Debye frequency \cite{BCS}.

The effective electric field ${\bf E}$ in Eq.~(\ref{KE}), as a gauge-invariant measurable quantity, is given by 
\begin{eqnarray}
e{\bf E}&=&-{\bm \nabla}_{\bf R}(e\phi+\mu_H+\mu_F)-\partial_te{\bf
  A}.\label{electric}
\end{eqnarray} 

The gauge-invariant density $n$ and current ${\bf j}$ read \cite{GOBE4}:
\begin{eqnarray}
en&=&e\sum_{\bf k}\left[1+{\rm Tr}(\rho^c_{\bf k}\tau_3)\right], \label{density}\\  
{\bf j}&=&\sum_{\bf k}{\rm
  Tr}\left(\frac{e{\bf k}}{m}\rho^c_{\bf k}\right).~~~\label{current}
\end{eqnarray}

We emphasize that Eq.~(\ref{KE}) is gauge-invariant under the gauge transformation first revealed by Nambu \cite{gi0}:
\begin{eqnarray}
eA_{\mu}&\rightarrow&eA_{\mu}-\partial_{\mu}\chi(R), \label{gaugestructure1}\\
\theta(R)&\rightarrow&\theta(R)+2\chi(R),\label{gaugestructure2}
\end{eqnarray}
where the four vectors $A_{\mu}=(\phi,{\bf A})$ and $\partial_{\mu}=(\partial_t,-{\bm \nabla}_{\bf R})$; $\theta$
denotes the phase of the superconducting order parameter. Thanks to the retained gauge invariance, the charge
conservation:
\begin{equation}
\partial_te\delta{n}+{\bf \nabla}_{\bf R}\cdot{\bf j}=0, \label{chargec}
\end{equation} 
is naturally satisfied during the electromagnetic
response as we proved in our latest work \cite{GOBE5}. This agrees 
with the Nambu's conclusion 
via the Ward's identity that the gauge invariance in the superconducting states is equivalent to the charge
conservation \cite{gi0}. Moreover, due to the gauge invariance, both the pump [third term in Eq.~(\ref{KE})] and drive
[sixth and seventh terms in Eq.~(\ref{KE})]
effects mentioned in the introduction are kept.

\subsection{Simplified GIKE in normal-skin-effect region}

In this part, we present a simplified GIKE in the normal-skin-effect region. 
We first choose a specific gauge by transforming Eq.~(\ref{KE}) under the gauge transformation $\rho_{\bf
  k}(R)={e^{-i\tau_3\theta(R)/2}}\rho^c_{\bf k}(R)e^{i\tau_3\theta(R)/2}$. Then, under a spatially uniform (i.e.,
long-wave-limit) optical field in the
normal-skin-effect region, the spatial gradient
terms in the kinetic equation can be neglected. Consequently, Eq.~(\ref{KE}) becomes:
\begin{eqnarray}
&&\partial_t\rho_{\bf
  k}\!+\!i\small[\small(\xi_k\!+\!\mu_{\rm
      eff}\small)\tau_3\!+\!|\Delta|\tau_1,\rho_{\bf 
k}\small]\!+\!\frac{i}{8}\small[|\Delta|\tau_1,\small({\bf p_s}\!\cdot\!\partial_{\bf k}\small)^2\rho_{\bf
k}\small]\nonumber\\
&&\mbox{}+\frac{1}{2}\left\{e{\bf E}\tau_3\!+\!{\bf p}_s|\Delta|\tau_2,\partial_{\bf k}\rho_{\bf
k}\right\}\!=\!\partial_t\rho_{\bf k}\Big|_{\rm
sc},
\label{GE}
\end{eqnarray}
with the gauge-invariant superconducting momentum ${\bf
  p}_s$ and effective field $\mu_{\rm eff}$ written as
\begin{eqnarray}
{\bf p}_s&=&\nabla_{\bf R}\theta-2e{\bf A},\label{ps}\\
\mu_{\rm eff}&=&\frac{\partial_t\theta}{2}+e\phi+\mu_{H}+\mu_{F}+\frac{p^2_s}{8m}.\label{ef}
\end{eqnarray}
Moreover, by expanding the density matrix as $\rho_{\bf k}=\sum^3_{i=0}\rho_{{\bf k}i}\tau_i$, the gap equations [Eqs.~(\ref{dp}) and (\ref{dm})] correspondingly read:
\begin{eqnarray}
g{\sum_{\bf k}}'\rho_{{\bf k}1}&=&-|\Delta|\label{gap},\\
g{\sum_{\bf k}}'\rho_{{\bf k}2}&=&0\label{phase}.
\end{eqnarray}
As shown in our latest work \cite{GOBE5},
Eq.~(\ref{gap}) gives the gap equation, from which one can self-consistently obtain the Higgs mode. The NG
mode can be self-consistently determined by Eq.~(\ref{phase}). 
Moreover, under the uniform optical response, one finds that ${\bm \nabla}_{\bf R}\cdot{\bf j}=0$. Therefore, as a consequence of the charge conservation [Eq.~(\ref{chargec})], 
the density fluctuation $\delta{n}$ and hence both the Hartree $\mu_H$ and Fock $\mu_F$ fields vanish.

\subsection{Microscopic Scattering}
\label{sesc}

We next present the scattering terms $\partial_t\rho_{\bf k}|_{\rm sc}$ in
Eq.~(\ref{GE}) which are derived based on the generalized Kadanoff-Baym
 ansatz \cite{spintronic,DS1,DS2,GKB}.  
Considering the fact that the electron-phonon
scattering is weak at low temperature, we mainly consider the
electron-impurity scattering.
The specific impurity scattering terms read (detailed derivation
can be found in Refs.~\onlinecite{DS1,spintronic,DS2}):
\begin{equation}
\partial_t\rho_{\bf k}\Big|_{\rm sc}=-[S_{\bf k}(>,<)-S_{\bf k}(<,>)+h.c.],\label{scat}
\end{equation}
with 
\begin{eqnarray}
S_{\bf k}(>,<)&=&n_i\sum_{\bf k'}\int^{t}_{-\infty}dt'[U_{\bf kk'}e^{i(t'-t)H_{\bf k'}}\rho^>_{\bf k'}(t')U_{\bf k'k}\nonumber\\
&&\mbox{}{\times}\rho^<_{\bf
k}(t')e^{-i(t'-t)H_{\bf k}}].\label{SK}
\end{eqnarray}
Here, $\rho^<_{\bf k}=\rho_{\bf
  k}$ and $\rho^>_{\bf k}=1-\rho_{\bf k}$; $H_{\bf k}=\xi_{{\bf k}+m{\bf v}_s\tau_3}\tau_3+|\Delta|\tau_1$ denotes the BCS
Hamiltonian in the presence of the superconducting velocity ${\bf v}_s$; $n_i$ is the impurity density;
$U_{\bf kk'}=V_{\bf k-k'}\tau_3$ stands for the electron-impurity interaction in the Nambu space.
This scattering term [Eq.~(\ref{scat})] is non-Markovian.

It is well established in semiconductor optics \cite{DS2} and
spintronics \cite{spintronic} that the clean-limit solution of the corresponding
kinetic (i.e., Liouville) equation:
\begin{equation}
\rho_{\bf k}^{>/<}(t')=e^{-i(t'-t)H_{\bf
      k}}\rho^{>/<}_{\bf k}(t)e^{i(t'-t)H_{\bf k}},\label{FCO}
\end{equation}
is substituted into the scattering terms as the Markovian approximation
to obtain the conventional energy conservation in the
scattering. In our previous works \cite{GOBE1,GOBE2,GOBE3,GOBE4},
we also take such approach in Eq.~(\ref{SK}) to derive 
the scattering in superconductors. In
the present work, this approach is sublated in the presence of the multi-cycle THz optical field, since the free
coherent oscillation in this circumstance does not hold, i.e., Eq.~(\ref{FCO}) is no longer the clean-limit solution of the GIKE in superconductors [Eq.~(\ref{GE})].  In fact, as 
shown in the next section, during
the multi-cycle THz pulse, the response of the
density matrix is forced to oscillate with the multiples of the optical frequency.

\section{Analytic Analysis}

In this section, by solving the simplified GIKE [Eq.~(\ref{GE})] in the normal-skin-effect region, we 
analytically investigate the scattering effect in the optical response of superconductors under multi-cycle THz pulse.
In this circumstance, analytic analyses for two extreme cases: during and after the pulse, are performed 
to carefully handle the Markovian approximation in order to turn the non-Markovian scattering in Eq.~(\ref{scat})
into the Markovian one.
The multi-cycle THz pulse,  as applied in recent experiments \cite{NL7}, possesses  a stable phase as well as a
narrow frequency bandwidth. Consequently, during the optical pulse, the system is under a periodic drive scheme at a well-defined
frequency,  similar to the case under a continuous waveform field.  In this situation,
the response of the superconductivity is forced to oscillate with
the multiples of the optical frequency.  
Whereas after the optical pulse, the system is free from the optical field, and the study in this situation reveals the relaxation
mechanism of the optically excited non-equilibrium states. 

\subsection{Forced oscillation}

During the multi-cycle THz pulse, 
by assuming the electromagnetic potential $\phi=\phi_0({\bf R})e^{i\omega{t}}$ and ${\bf A}={\bf
  A}_0e^{i\omega{t}}$, the density matrix $\rho_{\bf k}$ reads:
\begin{equation}
\rho_{\bf k}=\rho^0_{\bf k}+\rho^{\omega}_{\bf k}e^{i\omega{t}}+\rho^{2\omega}_{\bf k}e^{2i\omega{t}},\label{densitymatrix}
\end{equation} 
with the equilibrium-state density matrix $\rho^0_{\bf k}$ given by \cite{GOBE1,GOBE4,GOBE5}
\begin{equation}
\rho^0_{\bf k}=\frac{1}{2}-\frac{1-2f(E_k)}{2}\left(\frac{\xi_k}{E_k}\tau_3+\frac{\Delta_0}{E_k}\tau_1\right).\label{r0}
\end{equation}
Here, $\rho^{\omega (2\omega)}_{\bf  
  k}$ denotes the linear (second-order) response of the density matrix; $E_k=\sqrt{\xi_k^2+\Delta_0^2}$; $f(x)$ represents
the Fermi-distribution function.

Correspondingly, the responses of the phase $\theta$
and amplitude $|\Delta|$ of the superconducting order parameter are written as
\begin{eqnarray}
\theta&=&\theta^{\omega}e^{i\omega{t}}+\theta^{2\omega}e^{2i\omega{t}},\\
|\Delta|&=&\Delta_0+\delta|\Delta|^{\omega}e^{i\omega{t}}+\delta|\Delta|^{2\omega}e^{2i\omega{t}}.~~~~
\end{eqnarray} 
From Eqs.~(\ref{gap}) and (\ref{r0}), with
$g(E_k)=\frac{1-2f(E_k)}{2E_k}$, the equilibrium-state order parameter
$\Delta_0$ is determined by  
\begin{equation}
\Delta_0=-g{\sum_{\bf k}}'\rho^0_{{\bf k}1}=g{\sum_{\bf
    k}}'\left[\Delta_0g(E_k)\right],\label{gap0} 
\end{equation} 
which is exactly the gap
equation in the BCS theory \cite{BCS}.   
Moreover,
as shown in our latest work \cite{GOBE5}, the Higgs mode dose not manifest itself in the linear regime
($\delta|\Delta|^{\omega}=0$). 
The linear response of the NG mode from the GIKE \cite{GOBE5}, due to its coupling to the long-range Coulomb interaction, 
does not effectively occur either (i.e., $\mu_{\rm eff}^{\omega}=0$ and ${\bf p}^{\omega}_s=-2e{\bf A}_0^{\perp}$ with ${\bf A}_0^{\perp}$
being the physical transverse vector potential) as a result of the Anderson-Higgs mechanism \cite{AHM},  
in agreement with the previous works in the
literature \cite{AK,Apm,pm0,gi1,Ba0,Am1,gi0}.  

Furthermore, it is noted that in the presence of the multi-cycle THz pulse, 
the response of the density matrix [Eq.~(\ref{densitymatrix})], as the solution of the kinetic equation,  
is forced to
oscillate with the multiples of the optical frequency,
rather than the free coherent oscillation mentioned above. 
Then, substituting this forced oscillation [Eq.~(\ref{densitymatrix})] into the scattering term [Eq.~(\ref{scat})], the
$n$-th order of the scattering during the optical pulse 
can be obtained (refer to Appendix~\ref{aa}):
\begin{widetext}
\begin{eqnarray}
\partial_{t}\rho_{\bf k}|^{n\omega}_{\rm sc}&\!=\!&\!-\!n_i\pi\sum_{{\bf
    k'}\eta_1\eta_2}|V_{\bf k-k'}|^2[\tau_3\Gamma_{k'}^{\eta_1}(\tau_3\rho^{n\omega}_{\bf k}\!-\!\rho^{n\omega}_{\bf
  k'}\tau_3)\Gamma_{k}^{\eta_2}\delta(E_{\bf k'}^{\eta_1}\!+\!n\omega-E_{\bf k}^{\eta_2})\!+\!\Gamma_{k}^{\eta_2}(\rho^{n\omega}_{\bf
  k}\tau_3\!-\!\tau_3\rho^{n\omega}_{\bf
  k'})\Gamma_{k'}^{\eta_1}\tau_3\delta(E_{\bf k'}^{\eta_1}\!-\!n\omega\!-\!E_{\bf k}^{\eta_2})]\nonumber\\
&\!=\!&\!-\!n_i\pi\sum_{{\bf k'}}\sum^3_{i=0}|V_{\bf
  k-k'}|^2[Y^i_{\bf kk'}(n\omega)(\rho^{n\omega}_{{\bf
    k}i}\!-\!\rho^{n\omega}_{{\bf k'}i})\!+\!N^i_{\bf kk'}(n\omega)\rho^{n\omega}_{{\bf k'}i}],\label{nscat}
\end{eqnarray}\\
\end{widetext}
with
\begin{eqnarray}
Y^i_{\bf kk'}(n\omega)&=&\sum_{\eta_1\eta_2}(\tau_3\Gamma_{k'}^{\eta_1}\tau_3\tau_i\Gamma_{k}^{\eta_2}+\Gamma_{k}^{-\eta_2}\tau_i\tau_3\Gamma_{k'}^{-\eta_1}\tau_3)\nonumber\\
&&\mbox{}\times\delta(E_{\bf k'}^{\eta_1}+n\omega-E_{\bf k}^{\eta_2}),\label{Yi}\\
N^i_{\bf kk'}(n\omega)&=&\sum_{\eta_1\eta_2}(\tau_3\Gamma_{k'}^{\eta_1}[\tau_3,\tau_i]\Gamma_{k}^{\eta_2}+\Gamma_{k}^{-\eta_2}[\tau_i,\tau_3]\Gamma_{k'}^{-\eta_1}\tau_3)\nonumber\\
&&\mbox{}\times\delta(E_{\bf k'}^{\eta_1}+n\omega-E_{\bf k}^{\eta_2}).
\end{eqnarray}
Here, $\eta=\pm$; the projection operators
$\Gamma^{\pm}_{k}$ are written as {\small $\Gamma^{\pm}_{ 
    k}=U^{\dagger}_kQ^{\pm}U_k$} with {\small
  $Q^{\pm}=(1\pm\tau_3)/{2}$} and $U_k=u_k\tau_0-v_k\tau_++v_k\tau_-$  
being the unitary transformation matrix from the particle space to the
quasiparticle one. $u_k=\sqrt{1/2+\xi_k/(2E_k)}$ and
$v_k=\sqrt{1/2-\xi_k/(2E_k)}$; $E_{\bf k}^{\pm}={\bf k}\cdot{\bf v}_s\pm{E_k}$
denotes the tilted quasiparticle energy. It is noted that at low frequency
$\omega{\ll}\Delta_0$, the scattering term in Eq.~(\ref{nscat})
recovers the one in our previous work where we propose the three-fluid
model as mentioned in the introduction \cite{GOBE4}. In the present
work for the optical properties, we focus on the THz regime where
$\omega{\sim}\Delta_0$. Moreover, considering a weak and
fast-oscillating optical field, the tilt in quasiparticle energy
(i.e., Doppler shift ${\bf k}\cdot{\bf v}_s$), related to
electromagnetic field \cite{GOBE4}, can be neglected (i.e., $E_{\bf k}^{\pm}=\pm{E_k}$).

Then, as seen from
Eq.~(\ref{nscat}), due to the forced oscillation of the density matrix
by the influence of the multi-cycle THz pulse, the optical frequency
$\omega$ is involved in $\delta(E_{\bf k'}^{\eta_1}+n\omega-E_{\bf
      k}^{\eta_2})$ (i.e., the energy conservation of the
scattering). Consequently, besides the intraband
scattering ($\eta_1=\eta_2$), the interband
scattering channel ($\eta_1=-\eta_2$) is opened.

\subsubsection{Linear response: optical conductivity}
\label{LROC}

We first investigate the optical conductivity in the linear regime. 
The linear order of the GIKE [Eq.~(\ref{GE})] reads:
\begin{equation}
i\omega\rho^{\omega}_{\bf k}+i[\xi_{k}\tau_3+\Delta_0\tau_1,\rho^{\omega}_{\bf k}]+(e{\bf E}_0\cdot\partial_{\bf
  k})\rho^0_{{\bf k}3}\tau_0=\partial_{t}\rho_{\bf k}|^{\omega}_{\rm sc}.\label{LG}
\end{equation}
From above equation, it is noted that only the $\tau_0$ component of $\rho^{\omega}_{{\bf k}}$ 
is optically excited:
\begin{equation}
\rho^{\omega}_{{\bf k}0}=\rho^{\omega}_{{\bf k}0}|_{\rm cl}-\frac{n_i\pi}{i\omega}\sum_{{\bf k'}}|V_{\bf
  k-k'}|^2Y^0_{\bf kk'}(\omega)(\rho^{\omega}_{{\bf k}0}-\rho^{\omega}_{{\bf k'}0}),\label{r1}
\end{equation}
and the other components of $\rho^{\omega}_{{\bf
    k}}$ are zero, in consistency with the above mentioned vanishing $\delta|\Delta|^{\omega}$ [Eq.~(\ref{gap})] and
$\mu_{\rm eff}^{\omega}$ [Eq.~(\ref{phase})].
Here, {\small $\rho^{\omega}_{{\bf k}0}|_{\rm cl}=\frac{e{\bf
      E}_0\cdot{\bf k}}{im\omega}l(E_k)$} is the clean-limit solution 
with {\small  
  $l(E_k)=\partial_{\xi_k}[\xi_kg(E_k)]=-\frac{\Delta_0^2}{E_k}\partial_{E_k}g(E_k)-\partial_{E_k}f(E_k)$}  
consisting of superfluid
[$-\frac{\Delta_0^2}{E_k}\partial_{E_k}g(E_k)$] and quasiparticle [$-\partial_{E_k}f(E_k)$] contributions, 
exactly same as the one in
our previous works \cite{GOBE4,GOBE5}. The second term on the
right-hand side of Eq.~(\ref{r1}) comes from the scattering. 

The exact analytic solution of $\rho^{\omega}_{{\bf k}0}$ 
from Eq.~(\ref{r1}) is difficult in the presence of the scattering. Nevertheless, 
at the relatively weak scattering (i.e., $\xi<l$ with $\xi$ being the
coherence length), after the first-order iteration by substituting 
$\rho^{\omega}_{{\bf k}0}|_{\rm cl}$ into the
scattering term [second term on the right-hand side of Eq.~(\ref{r1})], $\rho^{\omega}_{{\bf k}0}$
can be directly solved:
\begin{equation}
\rho^{\omega}_{{\bf k}0}\approx\frac{e{\bf E}_0\cdot{\bf k}_F}{im\omega}l(E_k)+\frac{e{\bf E}_0\cdot{\bf
    k}_F}{m\omega^2}\eta_{\bf k},\label{sor1}
\end{equation}
with {\small $\eta_{\bf k}=n_i\pi\sum_{\bf k'}|V_{\bf k-k'}|^2Y^0_{\bf
    kk'}(\omega)[l(E_k)-\cos\theta_{\bf  kk'}l(E_{k'})]$}.
  
Then, substituting the solved $\rho^{\omega}_{{\bf k}0}$ into Eq.~(\ref{current}), the optical conductivity in the superconducting
states $\sigma_s(\omega)=\sigma_{1s}(\omega)+i\sigma_{2s}(\omega)$
is obtained (refer to Appendix~\ref{bb}):   
\begin{widetext}
\begin{eqnarray}
\frac{\sigma_{1s}(\omega)}{\sigma_{1n}(\omega)}&=&\Big\{\int^{\infty}_{\Delta_0}dE
\frac{[E(E+\omega)-\Delta^2_0][l(E)+l(E+\omega)]}{\sqrt{(E+\omega)^2-\Delta^2_0}\sqrt{E^2-\Delta^2_0}}-\int^{-\Delta_0}_{\Delta_0-\omega}dE\frac{[E(E+\omega)-\Delta^2_0]l(E+\omega)}{\sqrt{(E+\omega)^2-\Delta^2_0}\sqrt{E^2-\Delta^2_0}}\theta(\omega-2\Delta_0)\Big\},~~~~~\label{sigma1}\\
{\sigma_{2s}}(\omega)&=&-\frac{ne^2}{m\omega}+\sigma_{1n}(\omega)\int^{\Delta_0}_{{\rm max}(-\Delta_0,\Delta_0-\omega)}dE\frac{[E(E+\omega)-\Delta^2_0]l(E+\omega)}{\sqrt{(E+\omega)^2-\Delta^2_0}\sqrt{\Delta^2_0-E^2}}.\label{sigma2}
\end{eqnarray}\\
\end{widetext}
where $\theta(x)$ is the step function; 
{\small $\sigma_{1n}(\omega)=\frac{ne^2}{m\omega^2\tau_p}$} with
{\small $\frac{1}{\tau_p}=\Gamma_0-\Gamma_1$} exactly being the
momentum relaxation rate in normal metals and {\small
  $\Gamma_i=2n_i{\pi}D\int\frac{d\Omega_{\bf k'}}{4\pi}|V_{\bf 
  k_F-k_F'}|^2\cos{i}\theta_{\bf kk'}$}. $D$ is the density of states. It is noted that 
the first term in $\sigma_{2s}(\omega)$ recovers the clean-limit one in the superfluid as 
revealed in our previous work \cite{GOBE4}.  

Firstly, we point out that the obtained optical conductivity from the GIKE [Eqs.~(\ref{sigma1}) and~(\ref{sigma2})] becomes $\sigma_n=\frac{ne^2}{m\omega^2\tau_p}+\frac{ne^2}{im\omega}$
in the normal states at $T>T_c$ with $\Delta_0=0$ (refer to Appendix~\ref{rn}),
exactly recovering the one in normal metals as the Drude model or conventional Boltzmann equation revealed. To the best of our knowledge,
so far there is no theory of the optical conductivity in the literature
that can rigorously recover the conductivity in normal metals from $T<T_c$
to $T>T_{c}$, due to the difficulty in calculating the vertex
correction in superconductors \cite{NS,G1,vertex}. The GIKE here
actually provides an efficient approach to deal with the scattering.  

\begin{figure}[htb]
   {\includegraphics[width=8.1cm]{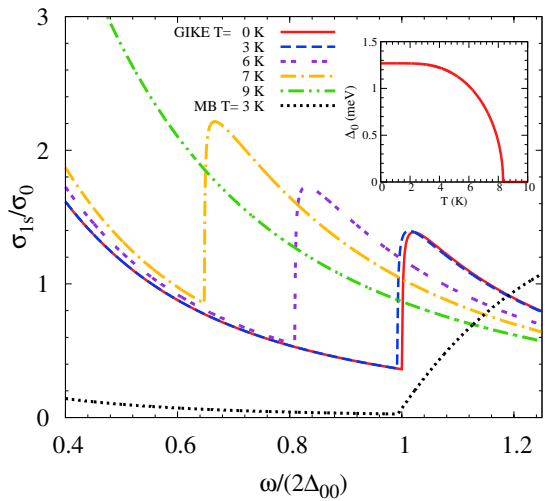}}
 \caption{(Color online) Frequency dependence of $\sigma_{1s}(\omega)$
   at different temperatures by calculating Eq.~(\ref{sigma1}).
   Constant $\sigma_0={n_0e^2}/{(me_0)}$. In our calculation,
   $\Delta_0$ is calculated from Eq.~(\ref{gap0});
   $\tau_p^{-1}=0.6\Delta_{00}$ with $\Delta_{00}$ denoting the order
   parameter at zero temperature; 
   $\sigma_{1n}(\omega)=\frac{\sigma_0e_0}{\omega^2\tau_p}$.  Other parameters used in our calculation are listed in
   Table~\ref{Table}. The black dotted curve denote the results from
   the MB theory, in which we artificially set
   $\sigma_{1n}(\omega)=\frac{4\sigma_0e_0}{\omega^2\tau_p}$ to
   enhance $\sigma_{1s}(\omega)$. 
 The inset shows the temperature dependence of the superconducting order parameter $\Delta_0$ to
   confirm the crossover point in the frequency dependence of 
   $\sigma_{1s}(\omega)$.}  
 \label{figyw1}
 \end{figure}

We then discuss the frequency dependence of the optical absorption
$\sigma_{1s}(\omega)$ in the superconducting states. In 
Eq.~(\ref{sigma1}), the first term originates from the intraband
scattering. Whereas the second one comes from the interband
scattering, leading to the step function. The frequency dependence
of $\sigma_{1s}(\omega)$ is plotted in Fig.~\ref{figyw1}.
As seen from the figure, $\sigma_{1s}(\omega)$ shows a significant crossover at
$\omega=2\Delta_0(T)$, which comes from the step function (i.e.,
    opened interband-scattering channel) for $\omega>2\Delta_0$ in Eq.~(\ref{sigma1}).
Secondly, at $T=0~$K with the finite superfluid contribution
$l(E_k)=\frac{\Delta_0^2}{2E_k^3}$ in Eq.~(\ref{sigma1}), one finds 
 that $\sigma_{1s}(\omega)$, shown by the solid curve
in Fig.~\ref{figyw1}, 
always exhibits a finite value even when $\omega<2\Delta_0$, in sharp
contrast to the vanishing 
$\sigma_{1s}(\omega)$ in the anomalous-skin-effect region as MB theory revealed. 
Moreover, as shown in Fig.~\ref{figyw1}, with the decrease of
$\omega$ from $\omega\gg2\Delta_0$, $\sigma_{1s}(\omega)$ first increases and then drops abruptly around $2\Delta_0$. By further decreasing $\omega$ below
$2\Delta_0$, due to the fast increase of $\sigma_{1n}(\omega)$ in Eq.~(\ref{sigma1}),
a significant upturn of $\sigma_{1s}(\omega)$
is observed. 

Results in the dirty limit ($l<\xi$) require a full numerical
calculation of Eq.~(\ref{r1}) and go beyond the 
analytic analysis. Nevertheless, from Eq.~(\ref{r1}), thanks to the
finite value of superfluid contribution in $l(E_k)$ at $T=0~$K, the finite
$\sigma_{1s}(\omega)$ at low temperature when $\omega<2\Delta_0$
is unlikely changed even in the dirty limit. In addition, 
due to the existence of $\delta(\omega-E_k+E_{k'})$ in $Y^0_{\bf
  kk'}(\omega)$ [Eq.~(\ref{Yi})], the crossover point at
$\omega=2\Delta_0(T)$ can also be obtained in the dirty limit.
These two points, by the full numerical calculation of Eq.~(\ref{r1}) in the dirty
limit, are justified (refer to Appendix~\ref{FNC}), in qualitative
agreement with the experimental findings
\cite{NSL1,NSL2,NSL10,NSL3,NSL4,NSL0,NSL11,NSL5,NSL6,NSL8,NSL9,NSL12}. 

As mentioned in the introduction, 
the MB theory derived from anomalous-skin-effect region is excessively used in the literature \cite{NSL1,NSL2,NSL4,NSL5,NSL6,NSL8,NSL9,NSL10,NSL11,NSL12}
to fit the experimental data in the normal-skin-effect region where the scattering effect
dominates. 
Nevertheless, as shown by the dotted curve in Fig.~\ref{figyw1}, at low temperature, when $\omega<2\Delta_0$,
$\sigma_{1s}(\omega)$ from MB theory derived at the anomalous-skin-effect region, which comes from the quasiparticle
contribution \cite{MB}, is too
small in comparison with the finite experimental observation \cite{NSL1,NSL2,NSL4,NSL5,NSL6,NSL8,NSL9,NSL10,NSL11,NSL12}.
Therefore, such an unphysical fit
underestimates the  
upturn of $\sigma_{1s}(\omega)$ below
$2\Delta_0$ particularly at low temperature, and hence, is incapable of 
capturing the experimental findings \cite{NSL1,NSL2,NSL3,NSL4,NSL5,NSL6,NSL8,NSL9,NSL10,NSL11,NSL12}.

\subsubsection{Second-order response: excitation of Higgs mode}
\label{SORH}

We next investigate the second-order response of the Higgs mode.
The second-order GIKE is written as
\begin{eqnarray}
&&2i\omega\rho^{2\omega}_{\bf k}\!+\!i[\xi_k\tau_3\!+\!\Delta_0\tau_1,\rho^{2\omega}_{\bf k}]\!+\!i[\mu^{2\omega}_{\rm
  eff}\tau_3\!+\!\delta|\Delta|^{2\omega}\tau_1,\rho^0_{\bf k}]\nonumber\\
&&\mbox{}\!+\!\frac{1}{2}\{e{\bf E}_0\tau_3\!+\!{\bf p}^{\omega}_s\tau_2\Delta_0,\partial_{\bf k}\rho^{\omega}_{\bf
  k}\}\!+\!\frac{i}{8}[\Delta_0\tau_1,({\bf p}^{\omega}_s\!\cdot\!{\partial_{\bf k}})^2\rho^0_{\bf k}]\nonumber\\
&&\mbox{}\!=\!\partial_{t}\rho_{\bf
  k}|^{2\omega}_{\rm sc},\label{G2}
\end{eqnarray} 
from which $\rho^{2\omega}_{\bf k}$ can be analytically solved at the relatively weak scattering. 

Substituting the solved $\rho^{2\omega}_{{\bf k}1}$ into Eq.~(\ref{gap}), the
second-order response of the Higgs mode can be self-consistently derived (refer to Appendix~\ref{cc}):
\begin{equation}
\delta|\Delta|^{2\omega}=\frac{\frac{v^2_F\Delta_0}{6}\Big(\frac{e{\bf
  E}_0}{i\omega}-\frac{{\bf p}^{\omega}_s}{2}\Big)^2d_\omega(1-is_H)}{\Delta_0^2-\omega^2+i\omega\gamma_H},\label{DS}
\end{equation}
where 
\begin{eqnarray}
&&d_{\omega}=\frac{\int^{\infty}_{\Delta_0}EdEo(E)d(E)}{\int^{\infty}_{\Delta_0}EdE\frac{g(E)o(E)}{{E^2-\Delta_0^2}}},\\
  &&\gamma_H=\frac{\Gamma_0{\rm F}[g]}{\int^{\infty}_{\Delta_0}EdE\frac{g(E)o(E)}{{E^2-\Delta_0^2}}},\\
&&s_H=\frac{\omega\Gamma_0{\rm F}[d]}{\int^{\infty}_{\Delta_0}EdEo(E)d(E)},
\end{eqnarray}
with {\small $d(E)=\frac{\partial_{E}l(E)}{E}$}, {\small
  $o(E)=\frac{\sqrt{E^2-\Delta^2_0}}{E^2-\omega^2}$} and
functional function
\begin{eqnarray}
&&{\rm F}[g]=\int^{\infty}_{\Delta_0}dEo(E){o(E\!+\!2\omega)}[g(E)\!+\!g(E\!+\!2\omega)]\nonumber\\
&&\mbox{}\!-\!\int^{2\omega\!-\!\Delta_0}_{\Delta_0}dE{o(E)}{o(2\omega\!-\!E)}{g}(2\omega\!-\!E)\theta(\omega\!-\!\Delta_0).~~~~   
\end{eqnarray}
It is noted that in the absence of the scattering (i.e., $\Gamma_0=0$), Eq.~(\ref{DS}) exactly reduces to the
clean-limit one revealed in our latest work \cite{GOBE5}.

\begin{figure}[htb]
  {\includegraphics[width=7.3cm]{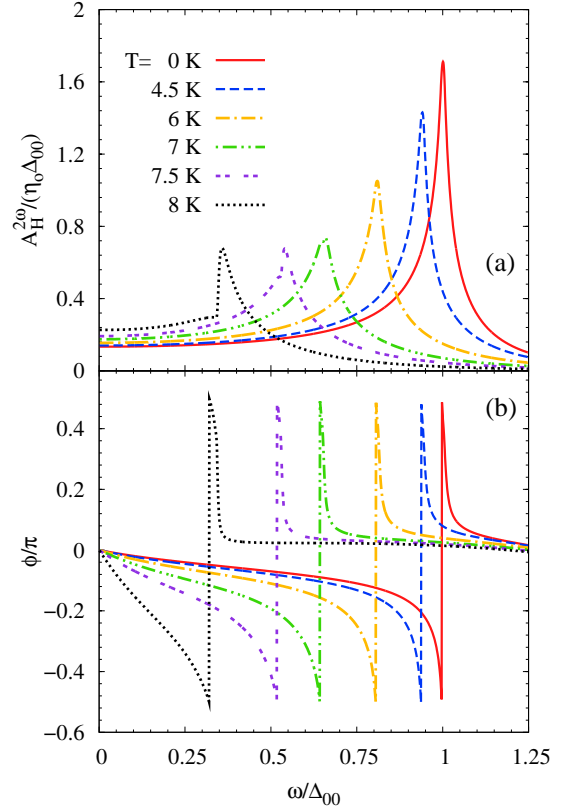}}
\caption{(Color online) Frequency dependence of the
magnitude $A^{2\omega}_H(\omega)$ and phase shift $\phi(\omega)$ of the second-order response of the Higgs mode 
at different temperatures. The dimensionless parameter
$\eta_0=\frac{v^2_F}{2\Delta^2_{00}}\Big(\frac{e{\bf
  E}_0}{i\omega}-\frac{{\bf p}^{\omega}_s}{2}\Big)^2$. $\Gamma_0=0.6\Delta_{00}$.  Other parameters used in our calculation are listed 
in Table~\ref{Table}.
}
  \label{figyw2}
\end{figure}

As seen from Eq.~(\ref{DS}), $\gamma_H$ from the scattering causes the broadening of the Higgs-mode spectrum
whereas $s_H$ represents the second-order optical absorption through the scattering. 
The existences of $s_H$ and $\gamma_H$ result in an imaginary part
in the second-order response of the Higgs mode, and hence, lead to a phase shift in this response. The magnitude $A^{2\omega}_H(\omega)$ and
phase shift $\phi(\omega)$ of the second-order response of the Higgs mode $\delta|\Delta|^{2\omega}=A_H^{2\omega}e^{i\phi(\omega)}$
are plotted in Fig.~\ref{figyw2}(a) and~(b), respectively. 
As seen from Fig.~\ref{figyw2}(a), 
the magnitude of the second-order response of
the Higgs mode exhibits a resonant peak at $2\omega=2\Delta_0(T)$, 
in consistency with the experimental observation \cite{NL2,NL3,NL4}. 
The phase shift $\phi$ of this second-order response
[Fig.~\ref{figyw2}(b)] exhibits a $\pi$-jump at $\omega=\Delta_0(T)$.
This is natural 
since from Eq.~(\ref{DS}),
the real part of
$\delta|\Delta|^{2\omega}$ at the weak scattering is proportional to $(\omega^2-\Delta^2_0)^{-1}$
whereas the imaginary one is proportional to $(\omega^2-\Delta^2_0)^{-2}$, leading to $\tan\phi\propto(\omega^2-\Delta^2_0)^{-1}$.  

\subsection{Free decay}

In the previous subsection, we have investigated the response of the superconducting states during the optical pulse. In this part, we
focus on the situation of the temporal evolution of the optically excited collective modes after the optical pulse.    

\subsubsection{Simplified model}
\label{Simmodel}

The GIKE after the optical pulse is written as
\begin{equation}
\partial_t\rho_{\bf
  k}\!+\!i\left[\left(\xi_k\!+\!\mu_{\rm
      eff}\right)\tau_3\!+\!|\Delta|\tau_1\!+\!\delta|\Delta|\tau_1,\rho_{\bf 
k}\right]\!=\!\partial_t\rho_{\bf k}\Big|_{\rm
sc}.
\label{FDKE}
\end{equation}
The density matrix is given by
\begin{equation}
\rho_{\bf k}=\rho^0_{\bf k}+\delta\rho_{\bf k},
\end{equation}
where $\delta\rho_{\bf k}$ denotes the part deviated from the equilibrium state due to the optical excitation. The fluctuations of the amplitude (i.e., $\delta|\Delta|$) and phase (i.e., $\mu_{\rm eff}$) of the order parameter can
be obtained from Eqs.~(\ref{gap}) and~(\ref{phase}), respectively.

It is noted that in Eq.~(\ref{FDKE}),
the second term on the left-hand side causes the coherent oscillation of the density matrix whereas the one on the right-hand side provides the scattering. 
In this circumstance, as established in the semiconductor optics \cite{DS2} and spintronics \cite{spintronic}, 
Eq.~(\ref{FCO}) as a clean-limit solution of Eq.~(\ref{FDKE}),
can be safely used into Eq.~(\ref{SK}) as the
Markov approximation to further derive the scattering terms. Then, the scattering 
which becomes free from the influence from the optical frequency,
is given by  
\begin{eqnarray}
\partial_t\rho_{\bf k}|_{\rm sc}&\!=\!&-n_i\pi\sum_{{\bf k'}\eta}|V_{\bf
  k\!-\!k'}|^2(\tau_3\Gamma^{\eta}_{k'}\tau_3\Gamma^{\eta}_{k}\rho_{\bf k}\!-\!\tau_3\rho_{\bf
  k'}\Gamma^{\eta}_{k'}\tau_3\Gamma^{\eta}_{k}\nonumber\\
&&\mbox{}\!+\!\rho_{\bf
  k}\Gamma^{\eta}_{k}\tau_3\Gamma^{\eta}_{k'}\tau_3\!-\!\Gamma^{\eta}_{k}\tau_3\Gamma^{\eta}_{k'}\rho_{\bf
  k'}\tau_3)\delta(E_{k'}\!-\!E_{k})\nonumber\\
&\!=\!&-n_i\pi\sum_{{\bf k'}}|V_{\bf
  k\!-\!k'}|^2[\tau_3(w_{k'k}\rho_{\bf k}\!-\!\rho_{\bf
  k'}w_{k'k})\!+\!(\rho_{\bf
  k}w_{kk'}\nonumber\\
&&\mbox{}\!-\!w_{kk'}\rho_{\bf
  k'})\tau_3]\delta(E_{k'}\!-\!E_{k}),\label{FDSC}
\end{eqnarray}
where
$w_{kk'}=\sum_{\eta}\Gamma^{\eta}_{k}\tau_3\Gamma^{\eta}_{k'}=w^1_{kk'}\tau_1+w^3_{kk'}\tau_3$
with $w^1_{kk'}=\frac{\Delta_0(\xi_{k'}+\xi_k)}{2E_kE_{k'}}$ and
$w^3_{kk'}=u_{k'}^2u_{k}^2+v^2_{k'}v^2_{k}-2u_kv_ku_{k'}v_{k'}$.

Since only the isotropic part of the density matrix
in the momentum space survives the summation in Eqs.~(\ref{gap}) and~(\ref{phase}),
i.e., contributes to the calculations of the amplitude and phase of the order parameter, we neglect the anisotropic part
in $\delta\rho_{\bf k}$. Then, considering the fact $w_{kk'}|_{\xi_k=-\xi_{k'}}=0$,
the scattering term in Eq.~(\ref{FDSC}) is simplified after the summation of ${\bf k'}$, and the GIKE becomes
\begin{eqnarray}
&&\partial_t\rho_{
  k}+i\left[\left(\xi_k+\mu_{\rm
      eff}\right)\tau_3+|\Delta|\tau_1+\delta|\Delta|\tau_1,\rho_{ 
k}\right]~~~~~\nonumber\\
&&\mbox{}=\!-\!2\Gamma_0{\rm sgn}(\xi_k)\Big[\rho_{k2}\frac{\xi_k}{E_k}\tau_2+\Big(\rho_{k1}\frac{\xi_k}{E_k}\!-\!\rho_{k3}\frac{\Delta_0}{E_k}\Big)\tau_1\Big].~~~~~\label{SFDKE}
\end{eqnarray}

Particularly, it is pointed out that Eq.~(\ref{SFDKE}) in the Anderson pseudospin picture \cite{As} is written as
\begin{equation}
\partial_{t}{\bf s}_k\!-\!2{\bf b}_k\times{\bf s}_k\!=\!-\!2\Gamma_0{\rm sgn}(\xi_k)\Big[({\bf s}_k\cdot{\bf a}_2){\bf {\hat x}}\!+\!\frac{\xi_k}{E_k}({
    \bf s}_k\cdot{\bf a}_1){\bf {\hat y}}\Big],\label{ASKE}
\end{equation}
where ${\bf b}_k=(\Delta_0+\delta\Delta,0,\xi_k+\mu_{\rm eff})$ and ${\bf s}_k=(\rho_{k1},\rho_{k2},\rho_{k3})$ denote the Anderson pseudo field
and spin, respectively; ${\bf a}_1=(0,1,0)$ and ${\bf a}_2=(\xi_k/E_k,0,-\Delta_0/E_k)$ are two transverse directions to
the equilibrium-state
pseudo field
${\bf b}^0_k$.  It is noted that in Eq.~(\ref{ASKE}), the second term on the left-hand side of the equation
causes the coherent precession
of the Anderson pseudospin, exactly same as the one in the previous works \cite{NL3,NL4,NL5,NL6,NL7,B1,B2,B3,B4,B5,B6}.
The terms on the right-hand side come from the scattering, which provide the relaxation of 
the non-equilibrium states.  Particularly, since $s^x_k$ and $s^y_k$ contribute to the calculations of the Higgs [Eq.~(\ref{gap})]
and NG [Eq.~(\ref{phase})] modes separately, 
one immediately finds that the first term on the right-hand side of Eq.~(\ref{ASKE}) provides the damping
of the excited Higgs mode whereas the second term causes the damping of the NG mode. 

We point out that in the present work, the relaxation terms on the right-hand side of Eq.~(\ref{ASKE}), exactly come
from the microscopic scattering, differing from and going beyond the previous phenomenological relaxation 
in the Anderson pseudospin picture mentioned in the introduction \cite{NL6,NL7}. 
In fact, the previous phenomenological relaxation mechanism,  
without the microscopic origin, is not exact
and convincing. Specifically, in Ref.~\onlinecite{NL6}, in analogy with the real spin precession,  the longitudinal and transverse relaxation processes, which describe 
the damping of the components of $\delta{\bf s}_k$ along and perpendicular ${\bf b}^0_k$,
are introduced into the Anderson pseudospin picture
through the phenomenological relaxation time.
Nevertheless, one finds that the longitudinal component of the 
pseudospin {\small $\delta{\bf s}_k\cdot{\bf b}^0_k/E_k=(\Delta_0\delta\rho_{k1}+\xi_k\delta\rho_{k3})/E_k=\delta\rho^q_{k3}$}. Since the diagonal {\small $\delta\rho^q_{k3}$} is related to
    the quasiparticle distribution, the longitudinal relaxation process [i.e., terms like ($\delta{\bf s}_{k}\cdot{\bf b}^0_{k}$)] directly describes the damping of the
    quasiparticles in which only the inelastic scattering contributes and the elastic scattering makes no contribution
    at all. Hence, in superconductors,
considering the weak inelastic electron-phonon scattering at low temperature,  the longitudinal relaxation process is
marginal and only the transverse ones  [i.e., terms like ($\delta{\bf s}_{k}\cdot{\bf a}_{1}$) and ($\delta{\bf s}_{k}\cdot{\bf a}_{2}$)] play the important role.
Particularly, there is no reason for
the two transverse relaxation processes, 
which provide the damping of the two
collective modes separately as mentioned above, to share the same rate. 
Most importantly, since ${\delta}s^z_{k}$ is related to the density fluctuation [i.e., $\delta{n}=\sum_{\bf k}{\delta}s^z_{k}=0$ from Eq.~(\ref{density})], as a consequence of the charge conservation,
the relaxation terms should not have any component along $z$ direction.
All above features, unsatisfied in Ref.~\onlinecite{NL6},  are well kept in our relaxation terms in Eq.~(\ref{ASKE})
, thanks to the microscopic scattering in the GIKE.

 \begin{figure}[htb]
   {\includegraphics[width=8.6cm]{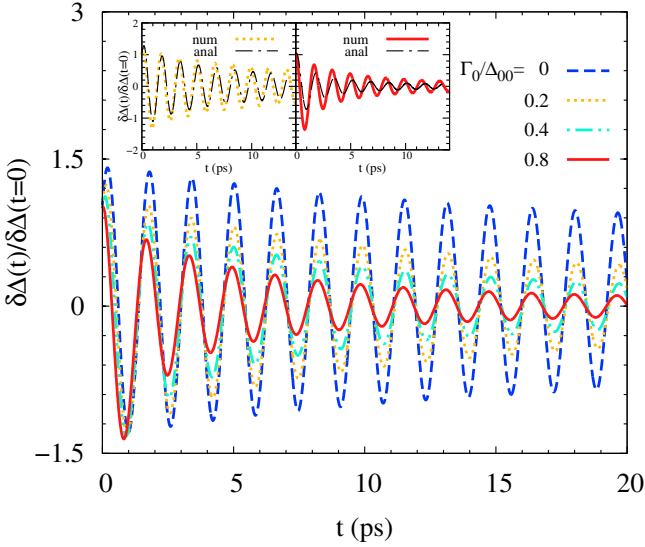}}
 \caption{(Color online) Temporal evolution of Higgs mode $\delta|\Delta|$ after the optical pulse at different scattering
strengths.  The inset shows the
comparison between the analytic solution from Eq.~(\ref{damping}) and full numerical results from
Eq.~(\ref{SFDKE}). In the
calculation, $\delta\rho_{\bf k}(t=0)=\rho^{\omega}_{\bf k}+\rho^{2\omega}_{\bf k}$ with $\omega=\Delta_0$ and $T=~1$K. 
Other parameters used in our calculation are listed in Table~\ref{Table}.}
  
 \label{figyw3}
 \end{figure}

\subsubsection{Damping of Higgs mode}
\label{DOHM}

By taking the optical response of the density matrix $\delta\rho_{\bf k}(t=0)=\rho^{\omega}_{\bf k}+\rho^{2\omega}_{\bf
  k}$, we first perform the numerical calculation to self-consistently solve Eq.~(\ref{SFDKE}) with Eqs.~(\ref{gap})
and~(\ref{phase}). Then, the temporal evolution of the Higgs mode  
$\delta|\Delta|(t)$ and NG mode $\mu_{\rm eff}(t)$ can be
self-consistently obtained. We focus on the measurable Higgs mode in
this part. 

The temporal evolution of the Higgs mode after the optical pulse is plotted in Fig.~\ref{figyw3} at different scattering
rates. As seen from the figure, $\delta|\Delta|(t)$ exhibits an oscillatory decay behavior, in consistency with the
experimental observation \cite{NL1,NL2,NL3,NL4,NL6,NL7}. 
The frequency of the
oscillation is around $2\Delta_0$, in agreement with the energy spectrum of the Higgs mode. Moreover, it is also found
that the damping of $\delta|\Delta|(t)$ shows a monotonic enhancement with the increase of the scattering rate.

To further understand the temporal evolution of $\delta|\Delta|(t)$, we analytically derive the solution of
Eq.~(\ref{SFDKE}) by first transforming 
Eq.~(\ref{SFDKE}) into the quasiparticle space through the unitary transformation $\rho^q_{\bf
  k}=U_k\rho_kU_k^{\dagger}$. Then, under a weak excitation (i.e., small $\delta\rho_{\bf k}$),
one has the components of the equation:
\begin{eqnarray}
&&\partial_t\delta\rho^q_{k+}+2iE_k\delta\rho^q_{k+}+\gamma_k\delta\rho^q_{k+}=2ia_k,~~~~\label{one}\\
&&\partial_t\delta\rho^q_{k-}-2iE_k\delta\rho^q_{k-}+\gamma_k\delta\rho^q_{k-}=-2ia_k,~~~~\label{two}\\
&&\partial_t\delta\rho^q_{k3}+(\delta\rho^q_{k+}+\delta\rho^q_{k-})\frac{\Delta_0}{E_k}\Gamma_0{\rm sgn}(\xi_k)=0,\label{three}
\end{eqnarray}
with  $\gamma_{k}=\Gamma_0{\rm sgn}(\xi_k)\frac{2\xi_k}{E_k}$ and {\small $a_k=(-\frac{\Delta_0}{E_k}\mu_{\rm eff}+\delta|\Delta|\frac{\xi_k}{E_k})\rho_{k3}^{q0}$}.

An exact solution from above equations is difficult. However, at the weak scattering, 
similar to the Elliot-Yafet mechanism in the spin relaxation of the semiconductor spintronics \cite{spintronic}, the coupling terms
between $\rho^q_{k3}$ and $\rho^q_{k\pm}$ in Eq.~(\ref{three}) 
can be {\em effectively} removed
through the unitary transformation 
as the L\"{o}wdin partition method showed \cite{dia}. Then,
$\rho^q_{\bf k}(t)$ and hence $\rho_{\bf k}(t)$ can be solved (refer to
Appendix~\ref{dd}). Consequently, from the gap equation [Eq.~(\ref{gap})], the temporal-evolution equation of the excited Higgs mode is given by: 
\begin{eqnarray}
&&\frac{\delta|\Delta|}{g}=\sum_{\bf
    k}\Big\{\frac{\Delta_0}{E_k}c_{k3}+\frac{\xi_k}{E_k}a_c\cos(2E_kt+\theta_c)e^{-\gamma_kt}\nonumber\\
&&\mbox{}-\left[\gamma_kg(E_k)\frac{\Delta_0^2}{E_k^2}\right]\int^t_0\delta|\Delta|(t')dt'+2E_kg(E_k)\frac{\xi_k^2}{E_k^2}\nonumber\\
&&\mbox{}\times\int^t_0\delta|\Delta|(t')\sin(2E_k\delta{t})e^{-\gamma_k\delta{t}}dt'\Big\},\label{damping}
\end{eqnarray}
with $a_{c}=\sqrt{c^2_1+c^2_2}$ and $\tan\theta_c=c_2/c_1$ and $\delta{t}=t-t'$. The coefficients $c_{ki}$ are determined by the initial optical excitation:
\begin{eqnarray}
c_{k1}&=&\frac{\Delta_0}{E_k}\delta\rho_{k3}(t=0)-\frac{\xi_k}{E_k}\delta\rho_{k1}(t=0),\\
c_{k2}&=&-\delta\rho_{k2}(t=0),\\
c_{k3}&=&-\frac{\Delta_0}{E_k}\delta\rho_{k1}(t=0)-\frac{\xi_k}{E_k}\delta\rho_{k3}(t=0).
\end{eqnarray}

As seen from the right-hand side of Eq.~(\ref{damping}), the first and
second terms are related with the initial excitation;
By only considering the third term, one has
{\small $\partial_t\delta|\Delta|=-\left[g\sum_{\bf k}\gamma_kg(E_k)\frac{\Delta_0^2}{E_k^2}\right]\delta|\Delta|$}. Thus, the third term on the right-hand side of Eq.~(\ref{damping}) causes the damping of $\delta|\Delta|$ with the relaxation rate proportional to $\gamma_k$.
The last term show the oscillatory
decay with the time evolution, and hence, directly lead to the oscillating damping of $\delta|\Delta|$ with the relaxation rate
proportional to $\gamma_k$. The relaxation rate of the Higgs mode
therefore increases by increasing the impurity density,
similar to the Elliot-Yafet mechanism in the spin relaxation of the semiconductor spintronics \cite{spintronic}. 
Comparisons between the analytic solution [Eq.~(\ref{damping})] and
full numerical results are plotted in the insets of Fig.~\ref{figyw3},
where the results from the two sets of calculations agree well
with each other.

Finally, from Eq.~(\ref{damping}), it is found that the
long-time dynamic of the Higgs mode behaves as (refer to Appendix~\ref{DFHD})
\begin{equation}
      \label{FHD}
      \delta\Delta(t)\sim\frac{\cos(2\Delta_0t)e^{-{\bar \gamma}t}}{\sqrt{\Delta_0t}},
\end{equation}
exhibiting an oscillatory decay behavior with oscillating frequency at
the Higgs-mode energy $2\Delta_0$. Here, ${\bar \gamma}$ is the
average of ${\gamma_{\bf k}}$ in the momentum space. 
In the absence of disorder
(${\bar \gamma}=0$), Eq.~(\ref{FHD}) reduces to the previous coherent BCS
oscillatory decay \cite{B2,B3,OD1,OD2,OD3,OD4} as it should be, since our kinetic equation
[Eq.~(\ref{ASKE}) or Eq.~(\ref{SFDKE})] without the scattering exactly
recovers the linearized Bloch (i.e., Anderson-pseudospin) equations around
the equilibrium state \cite{B2,B3,OD3,OD4}. Whereas the presence of the impurity leads to
exponential decay.

\section{SUMMARY AND DISCUSSION}
\label{summary}

Within the GIKE approach, we analytically investigate the influence of
the scattering on the optical response of 
superconductors in the normal-skin-effect region ({\small $l<\delta$}). 
Two extreme situations: during and after 
a multi-cycle THz pulse pulse, are considered
with a careful implementation 
of the Markovian approximation for the microscopic
scattering.  During the pulse,  the multi-cycle optical field with the stable phase and 
narrow frequency bandwidth as applied in recent experiments \cite{NL7},
exhibits the continuous-wave-like behavior.  Then, response of the density of
matrix, as the solution of the free GIKE in superconductors, is forced
to oscillate with the multiples of the optical frequency.  Consequently, 
due to this forced oscillation, after the Markovian
approximation, the energy conservation of the scattering is
influenced by the optical frequency. 
Whereas after the optical pulse, the 
system is free from the optical field, and the density of matrix in
this situation exhibits the free coherent oscillation in the clean limit. 
Then, after the Markovian approximation,
the energy conservation of the scattering becomes free from the
influence from the optical frequency. 
Rich physics in both extreme cases is revealed. 

Specifically, during the pulse, responses of the superconductivity in
linear and second-order regimes are studied.  
In the linear regime, we analytically derive the optical conductivity
from the GIKE at the weak scattering ({\small 
  $l>\xi$}). We show that by taking $T>T_c$
the optical conductivity from our theory obtained at $T<T_c$ exactly recovers the one
in normal metals as the Drude model or conventional Boltzmann equation revealed. 
To the best of our knowledge, so far there is no theory in the literature
that can rigorously make this recovery. 
Whereas in the superconducting states, 
we find that the optical absorption $\sigma_{1s}(\omega)$,
due to the contribution of superfluid
density,  always exhibits a finite value when
$\omega<2\Delta_0$ even at low temperature, and shows an upturn
with the decrease of frequency below $2\Delta_0$,
in contrast to the vanishing $\sigma_{1s}(\omega)$ in
the anomalous-skin-effect region as MB theory revealed \cite{MB}. 
Moreover, $\sigma_{1s}(\omega)$ shows a significant crossover at
$\omega=2\Delta_0(T)$, which comes from opened interband-scattering
channel for $\omega>2\Delta_0$.
Through the full numerical calculation, we
further show that  
both the upturn of the finite $\sigma_{1s}(\omega)$ below $2\Delta_0$
and the crossover point at $\omega=2\Delta_0(T)$ in
$\sigma_{1s}(\omega)$ also appear in the dirty-limit regime ({\small
  $\xi<l$}), in qualitative agreement with the experimental observations in 
disordered type-II superconductors like Nb \cite{NSL1,NSL2}, NbN
\cite{NSL10}, MgB$_2$ \cite{NSL3,NSL4,NSL0,NSL11}, NbTiN \cite{NSL5,NSL6,NSL8}
and Al \cite{NSL9,NSL12}.

As for the second-order regime, we study the
response of the Higgs mode. We show that 
the scattering causes a phase shift in this second-order optical response. 
Particularly, we find that this
phase shift exhibits a significant $\pi$-jump at $\omega=\Delta_0$, which provides a very clear feature for the
experimental detection. Recently, thanks to the advanced pump-probe
technique, a $\pi$-jump of the phase shift has been experimentally observed at 
$\omega=\Delta_0(T)$ in the second-order optical response of the disordered high-$T_c$ cuprates-based superconductors \cite{NL7}. The origin of this jump is still
controversial. Whereas our present work suggests 
that the $\pi$-jump of the phase shift in the second-order
optical response can also be realized in the conventional
superconductors through the scattering effect.

Finally, we study the relaxation mechanism of the excited collective
modes after the pulse. In this situation,  
based on the complete GIKE, a simplified model with the damping terms
in the Anderson pseudospin picture 
 is proposed. The damping terms in this model exactly come from the microscopic scattering,
 differing from and going beyond the phenomenological relaxation
 mechanism in the previous works \cite{NL6,NL7}. 
 Particularly, both the charge conservation and 
the unique feature of the dominant elastic scattering in
superconductors: vanishing longitudinal relaxation process, are 
kept in our relaxation terms, in sharp contrast to Ref.~\onlinecite{NL6}. 
Then, by studying the damping of the Higgs-mode 
excitation, we reveal an exponential relaxation mechanism due
to the elastic scattering, which 
shows a monotonic enhancement with the increase of the impurity
density. In addition, we also investigate the damping of the NG mode
(refer to Appendix~\ref{ee}). 
It is found that in the conventional BCS superconductors, the damping
of the phase fluctuation (NG mode) is much faster than that of the
amplitude fluctuation (Higgs mode) of the order parameter.

{\em Note added}: After the completion of our manuscript, we became
aware of a very recent paper by Silaev\cite{CO}. In that paper, by separately using Eilenberger equation and diagram 
formalism, the author studied the Higgs mode excitation in the
presence of the scattering. This is indeed the very first paper that
rigorously calculates the scattering influence on optical properties  
within the Eilenberger equation in the
  literature, even though it is too complex to
  obtain final analytic solution. Nevertheless, based on the following
  reasons, the results in that paper are not correct.  
  Firstly, in Ref.~\onlinecite{CO} by Silaev, the
  claimed conclusion that the Higgs-mode generation is zero without impurity is based on the incomplete electromagnetic effect in his approach. Specifically, both the Hamiltonian used in his 
  diagram formalism and the Eilenberger equation are not
  gauge invariant with vector potential ${\bf A}$ alone \cite{GEG}. 
  It is well known that the gauge invariance is the basic
  character of the electromagnetic field. The absence of the 
  gauge invariance indicates that the incomplete electromagnetic effect.
  Secondly, another
  conclusion in
  Ref.~\onlinecite{CO}
  that the Higgs mode is not 
  sensitive to disorder, is also incorrect. 
  This can be easily
  seen by the following simple analysis through the general
  physics.  In the Nambu space, the BdG Hamiltonian in the
  presence of the Higgs mode excitation is written as $H_{\rm
    BdG}=\xi_{\hat p}\tau_3+\Delta_0({\bf r})\tau_1+\delta|\Delta({\bf
    r})|\tau_1$ in the real space, and the electron-impurity
  interaction is given by $V({\bf r})\tau_3$. Then, due to the
      non-commutation relation
      \begin{equation}
        [\delta|\Delta({\bf r})|\tau_1, V({\bf r})\tau_3]\ne0,\label{RE}
        \end{equation}
  the Higgs mode must be sensitive to the disorder. 
  In fact, the scattering influence on the Higgs mode in the present work
  exactly comes from this non-commutation relation. Specifically, 
  our scattering term of the isotropic part [Eq.~(\ref{FDSC}) with
  $w_{kk'}|_{\xi_k=-\xi_{k'}}=0$] is given by 
  \begin{equation}
    \partial_{t}\rho_k|_{sc}\!=\!\!-\!\frac{\Gamma_0}{2}\!\int\!{d\xi_{k'}}\Big(\tau_3\Big[\small\sum_{\eta}\Gamma^{\eta}_k\tau_3\Gamma^{\eta}_k,\delta\rho_{k}\Big]+h.c.\Big)\delta(E_k\!-\!E_{k'}), 
  \end{equation} 
  in which the projection operator $\Gamma^{\eta}_k$ picks up the
  energy-conserved scattering channel. Then, it is immediately observed
  that the Higgs-mode part ($\tau_1$ component of $\delta\rho_{k}$)
  $[\sum_{\eta}\Gamma^{\eta}\tau_3\Gamma^{\eta},\delta\rho_{k1}\tau_1]$
  has the form of Eq.~(\ref{RE}) limited by the energy conservation.
  
\begin{table}[htb]
  \caption{The used parameters in our calculations. 
 With the specific values of $\Delta_{00}$ and $\omega_D$,
the effective electron-electron attractive potential $g$ is
    determined by Eq.~(\ref{gap0}) at $T=0~$K.} 
\label{Table}
  \begin{tabular}{l l l l}
    \hline
    \hline   
    $\Delta_{00}$&\;\;\;\;$1.268~$meV&\;\;\;$\omega_D$&\;\;\;\;$15.856~$meV\\
    $E_F$&\;\;\;\;$700~$meV&\;\;\;$e_0$&\;\;\;\;$8~$meV\\
    $(eE_0/i\omega)^2/m$&\;\;\;\;$10^{-4}\Delta_{00}$&\;\;\;$A^{\perp}_0$&\;\;\;\;$0$\\   
    \hline
    \hline
\end{tabular}
\end{table}  

\begin{acknowledgments}
This work was supported by the National Natural Science Foundation of 
China under Grants No.\ 11334014 and No.\ 61411136001.  
\end{acknowledgments}

\begin{appendix}

\section{Derivation of Eq.~(\ref{nscat})}
\label{aa}

In this part, we derive Eq.~(\ref{nscat}). From Eq.~(\ref{SK}), one has
\begin{widetext}
\begin{eqnarray}
&&I_{\bf k}=S_{\bf k}(>,<)-S_{\bf k}(<,>)=n_i\sum_{{\bf k'}\eta_1\eta_2}\int^{t}_{-\infty}dt'|V_{\bf k\!-\!k'}|^2{e^{-i(t-t')(E^{\eta_1}_{\bf k'}-E_{\bf k}^{\eta_2})}}\{\tau_3\Gamma^{\eta_1}_{k'}[\tau_3\rho_{\bf
  k}(t')-\rho_{\bf
  k'}(t')\tau_3]\Gamma^{\eta_2}_k\},
\end{eqnarray}
\end{widetext}
in which $e^{itH_{\bf
    k}}=\sum_{\eta}\Gamma^{\eta}_{k}e^{itE^{\eta}_{\bf k}}$ is used.

The $n$-th order of above equation during the optical response
is written as
\begin{widetext}
\begin{eqnarray}
I_{\bf k}|^{n\omega}&=&{n_i}\!\sum_{{\bf k'}\eta_1\eta_2}\!|V_{\bf k\!-\!k'}|^2[\tau_3\Gamma^{\eta_1}_{k'}(\tau_3\rho^{n\omega}_{\bf
  k}\!-\!\rho^{n\omega}_{\bf
  k'}\tau_3)\Gamma^{\eta_2}_k]\int^{0}_{-\infty}dt'{e^{i(E^{\eta_1}_{\bf k'}-E_{\bf k}^{\eta_2}+n\omega)t'}}=\mbox{}{n_i}\!\sum_{{\bf k'}\eta_1\eta_2}\!|V_{\bf k\!-\!k'}|^2\frac{\tau_3\Gamma^{\eta_1}_{k'}(\tau_3\rho^{n\omega}_{\bf
  k}\!-\!\rho^{n\omega}_{\bf
  k'}\tau_3)\Gamma^{\eta_2}_k}{i(E^{\eta_1}_{\bf k'}\!-\!E_{\bf k}^{\eta_2}\!+\!n\omega-i0^+)}\nonumber\\
&=&\mbox{}\pi{n_i}\!\sum_{{\bf k'}\eta_1\eta_2}\!|V_{\bf k\!-\!k'}|^2[\tau_3\Gamma^{\eta_1}_{k'}(\tau_3\rho^{n\omega}_{\bf
  k}-\rho^{n\omega}_{\bf
  k'}\tau_3)\Gamma^{\eta_2}_k]\delta(E^{\eta_1}_{\bf k'}-E_{\bf
  k}^{\eta_2}+n\omega).
\end{eqnarray}
\end{widetext}
Similarly, one also finds
\begin{eqnarray}
I^{\dagger}_{\bf k}|^{n\omega}&=&\pi{n_i}\sum_{{\bf k'}\eta_1\eta_2}|V_{\bf k\!-\!k'}|^2[\Gamma_{k}^{\eta_2}(\rho^{n\omega}_{\bf
  k}\tau_3\!-\!\tau_3\rho^{n\omega}_{\bf
  k'})\Gamma_{k'}^{\eta_1}\tau_3]\nonumber\\
&&\mbox{}\times\delta(E^{\eta_1}_{\bf k'}-E_{\bf
  k}^{\eta_2}-n\omega).
\end{eqnarray}
Consequently, Eq.~(\ref{nscat}) is derived. For completeness, 
The explicit expressions of $Y^i_{\bf kk'}(n\omega)$ [Eq.~(\ref{Yi})] are given by
\begin{widetext}
\begin{eqnarray}
Y^0_{\bf
  kk'}&=&\tau_0[(u^2_ku^2_{k'}+v^2_kv^2_{k'}-2u_ku_{k'}v_kv_{k'})\delta(E_{k'}+n\omega-E_k)+(u^2_ku^2_{k'}+v^2_kv^2_{k'}-2u_ku_{k'}v_kv_{k'})\delta(E_{k}+n\omega-E_{k'})\nonumber\\
&&\mbox{}+(u^2_kv^2_{k'}+v^2_ku^2_{k'}+2u_ku_{k'}v_kv_{k'})\delta(E_{k'}+n\omega+E_k)+(u^2_kv^2_{k'}+v^2_ku^2_{k'}+2u_ku_{k'}v_kv_{k'})\delta(n\omega-E_k-E_{k'})],~~~~~\label{Y0}\\
Y^3_{\bf
  kk'}&=&[(u^2_ku^2_{k'}+v^2_kv^2_{k'}+2u_ku_{k'}v_kv_{k'})\tau_3+\Delta_0(\xi_{k'}-\xi_k)/(2E_kE_{k'})\tau_1+i(u_kv_k+u_{k'}v_{k'})\tau_2]\delta(E_{k'}+n\omega-E_k)\nonumber\\
&&\mbox{}+[(u^2_ku^2_{k'}+v^2_kv^2_{k'}+2u_ku_{k'}v_kv_{k'})\tau_3+\Delta_0(\xi_{k'}-\xi_k)/(2E_kE_{k'})\tau_1-i(u_kv_k+u_{k'}v_{k'})\tau_2]\delta(E_{k}+n\omega-E_{k'})\nonumber\\
&&\mbox{}+[(u^2_kv^2_{k'}+v^2_ku^2_{k'}-2u_ku_{k'}v_kv_{k'})\tau_3-\Delta_0(\xi_{k'}-\xi_k)/(2E_kE_{k'})\tau_1+i(u_{k'}v_{k'}-u_kv_k)\tau_2]\delta(E_{k}+n\omega+E_{k'})\nonumber\\
&&\mbox{}+[(u^2_kv^2_{k'}+v^2_ku^2_{k'}-2u_ku_{k'}v_kv_{k'})\tau_3-\Delta_0(\xi_{k'}-\xi_k)/(2E_kE_{k'})\tau_1+i(u_{k}v_{k}-u_{k'}v_{k'})\tau_2]\delta(n\omega-E_{k'}-E_k),~~~~~~\\
Y^1_{\bf
  kk'}&=&[\Delta_0(\xi_{k'}-\xi_k)/(2E_kE_{k'})\tau_3+(u^2_kv^2_{k'}+v^2_ku^2_{k'}-2u_ku_{k'}v_kv_{k'})\tau_1+i(u^2_{k'}v_k^2-u^2_kv^2_{k'})\tau_2]\delta(E_{k'}+n\omega-E_k)\nonumber\\
&&\mbox{}+[\Delta_0(\xi_{k'}-\xi_k)/(2E_kE_{k'})\tau_3+(u^2_kv^2_{k'}+v^2_ku^2_{k'}-2u_ku_{k'}v_kv_{k'})\tau_1-i(u^2_{k'}v_k^2-u^2_kv^2_{k'})\tau_2]\delta(E_{k}+n\omega-E_{k'})\nonumber\\
&&\mbox{}+[\Delta_0(\xi_{k}-\xi_{k'})/(2E_kE_{k'})\tau_3+(u^2_ku^2_{k'}+v^2_kv^2_{k'}+2u_ku_{k'}v_kv_{k'})\tau_1+i(u^2_{k'}u_k^2-v^2_kv^2_{k'})\tau_2]\delta(E_{k}+n\omega+E_{k'})\nonumber\\
&&\mbox{}+[\Delta_0(\xi_{k}-\xi_{k'})/(2E_kE_{k'})\tau_3+(u^2_ku^2_{k'}+v^2_kv^2_{k'}+2u_ku_{k'}v_kv_{k'})\tau_1-i(u^2_{k'}u_k^2-v^2_kv^2_{k'})\tau_2]\delta(n\omega-E_k-E_{k'}),\\
iY^2_{\bf
  kk'}&=&[(u_{k}v_{k}+u_{k'}v_{k'})\tau_3+(u^2_{k'}v_k^2-u^2_kv^2_{k'})\tau_1+i(u^2_kv^2_{k'}+v^2_ku^2_{k'}+2u_ku_{k'}v_kv_{k'})\tau_2]\delta(E_{k'}+n\omega-E_k)\nonumber\\
&&\mbox{}-[(u_{k}v_{k}+u_{k'}v_{k'})\tau_3+(u^2_{k'}v_k^2-u^2_kv^2_{k'})\tau_1-i(u^2_kv^2_{k'}+v^2_ku^2_{k'}+2u_ku_{k'}v_kv_{k'})\tau_2]\delta(E_{k}+n\omega-E_{k'})\nonumber\\
&&\mbox{}+[(u_{k'}v_{k'}-u_{k}v_{k})\tau_3+(u^2_{k'}u_k^2-v^2_kv^2_{k'})\tau_1+i(u^2_ku^2_{k'}+v^2_kv^2_{k'}-2u_ku_{k'}v_kv_{k'})\tau_2]\delta(E_{k'}+n\omega+E_k)\nonumber\\
&&\mbox{}-[(u_{k'}v_{k'}-u_{k}v_{k})\tau_3+(u^2_{k'}u_k^2-v^2_kv^2_{k'})\tau_1-i(u^2_ku^2_{k'}+v^2_kv^2_{k'}-2u_ku_{k'}v_kv_{k'})\tau_2]\delta(n\omega-E_k-E_{k'}).\label{Y2}
\end{eqnarray}
\end{widetext}
One also has
{\small $N^{1}_{\bf kk'}(n\omega)=2Y^1_{\bf
  kk'}(n\omega)$}, {\small $N^{2}_{\bf kk'}(n\omega)=2Y^2_{\bf kk'}(n\omega)$} and {\small $N^0_{\bf kk'}(n\omega)=N^3_{\bf kk'}(n\omega)=0$}.

\section{Derivation of Eqs.~(\ref{sigma1}) and (\ref{sigma2})}
\label{bb}

We derive Eqs.~(\ref{sigma1}) and (\ref{sigma2}) in this part.
At the weak scattering, substituting the solved $\rho^{\omega}_{{\bf
    k}0}$ [Eq.~(\ref{sor1})] into Eq.~(\ref{current}), one has
{\small
\begin{equation}
{\bf j}={\bf j}_1+{\bf j}_2,
\end{equation}
with
\begin{eqnarray}
{\bf j}_1&=&\frac{2e^2{\bf E}_0Dk_F^2}{3i\omega{m^2}}\int{d\xi_k}l(E_k)=\frac{ne^2}{i\omega{m}}{\bf E}_0,\label{j1}\\
{\bf j}_2&\approx&\sum_{\bf kk'}\frac{ne^2{\bf
    E}_0}{m\omega^2}n_i\pi|V_{\bf
  k_Fk'_F}|^2Y^0_{\bf kk'}(\omega)[l(E_k)\!-\!\cos\theta_{{\bf kk'}}{l(E_{k'})}]\nonumber\\
&&\mbox{}\nonumber\\
&=&\sum_{\bf k}\frac{ne^2{\bf E}_0}{2m\omega^2}\!\int\!{d\xi_{k'}}Y^0_{\bf kk'}(\omega)[\Gamma_0l(E_k)\!-\!\Gamma_1l(E_{k'})].~~~~~~\label{j2}
\end{eqnarray}}

Then, with the explicit expression of $Y^0_{\bf kk'}$ in Eq.~(\ref{Y0}), 
the above equation becomes
\begin{eqnarray}
{\bf j}_2&=&\frac{ne^2{\bf
    E}_0}{m\omega^2}\frac{1}{2}\int{d\xi_k}\int{d\xi_{k'}}[\Gamma_0l(E_k)-\Gamma_1l(E_{k'})]\nonumber\\
&&\mbox{}\times\sum_{\eta_1\eta_2}\frac{1}{2}\Big(1+\frac{\Delta_0}{E^{\eta_1}_kE^{\eta_2}_{k'}}\Big)\delta(\omega+E^{\eta_1}_{k}+E^{\eta_2}_{k'})
]\nonumber\\
&=&\frac{ne^2{\bf
    E}_0}{m\omega^2}\int{dE}\int{dE'}\frac{EE'[\Gamma_0l(E)-\Gamma_1l(E')]}{\sqrt{E^2-\Delta^2_0}\sqrt{{E'}^2-\Delta^2_0}}\nonumber\\
&&\mbox{}\times\sum_{\eta_1\eta_2}\Big(1+\frac{\eta_1\eta_2\Delta_0}{EE'}\Big)\delta(\omega+\eta_1E+\eta_2E'),
\end{eqnarray}
in which we have taken care of
the particle-hole symmetry to remove terms with the
odd orders of $\xi_k$ and $\xi_{k'}$ in the summation of ${\bf k}$ and ${\bf k'}$.
After the mathematical integral, above equation becomes
{\small
\begin{eqnarray}
&&{\bf j}_2=\sigma_{1n}{\bf
  E}_0\Big\{\int^{\infty}_{\Delta_0}dE\frac{[(E\!+\!\omega)E\!-\!\Delta^2_0][l(E\!+\!\omega)\!+\!l(E)]}{\sqrt{E^2\!-\!\Delta^2_0}\sqrt{(E\!+\!\omega)^2\!-\!\Delta^2_0}}\nonumber\\
&&\mbox{}+\int^{\omega\!-\!\Delta_0}_{\Delta_0}dE\frac{[(\omega\!-\!E)E\!+\!\Delta^2_0][l(\omega\!-\!E)\!+\!l(E)]}{2\sqrt{E^2\!-\!\Delta^2_0}\sqrt{(\omega\!-\!E)^2\!-\!\Delta^2_0}}\theta(\omega\!-\!2\Delta_0)\nonumber\\
&&\mbox{}+\frac{1}{i}\int^{\omega\!+\!\Delta_0}_{{\rm max}(\omega\!-\!\Delta_0,\Delta_0)}dE\frac{[(\omega\!-\!E)E\!+\!\Delta^2_0][l(\omega\!-\!E)\!+\!l(E)]}{2\sqrt{E^2\!-\!\Delta^2_0}\sqrt{\Delta^2_0\!-\!(\omega\!-\!E)^2}}\Big\}.\nonumber\\
\end{eqnarray}
} 

Consequently, the optically excited current ${\bf j}$ in the linear regime and hence the optical conductivity
$\sigma_{s}(\omega)=\sigma_{1s}(\omega)+i\sigma_{2s}(\omega)$ are derived. 

\section{Optical conductivity at $T>T_c$}
\label{rn}

We give the optical conductivity at $T>T_c$. In the normal state at $T>T_c$, with $\Delta_0=0$, one finds that
{\small $l(E)=-\partial_Ef(E)$} and {\small
  $\frac{E(E+\omega)-\Delta^2_0}{\sqrt{(E+\omega)^2-\Delta^2_0}\sqrt{E^2-\Delta^2_0}}=E(E+\omega)/(|E||E+\omega|)$}. 
Then, thanks to the constant density of states in normal states,
Eqs.~(\ref{sigma1}) and (\ref{sigma2}) become
\begin{eqnarray}
&&\frac{\sigma_{1s}(\omega)}{\sigma_{1n}(\omega)}=\!-\!\int^{\infty}_{0}dE[\partial_Ef(E)\!+\!\partial_{E\!+\!\omega}f(E\!+\!\omega)]\!-\!\int^{0}_{-\omega}dE\nonumber\\
&&\mbox{}\times\partial_{E+\omega}f(E+\omega)=-2\int^{\infty}_{0}dE\partial_Ef(E)=1,\\
&&{\sigma_{2s}}(\omega)=-\frac{ne^2}{m\omega},
\end{eqnarray}
which are exactly the optical conductivity in normal metals as the Drude model or conventional Boltzmann equation revealed.

 \begin{figure}[htb]
   {\includegraphics[width=7.8cm]{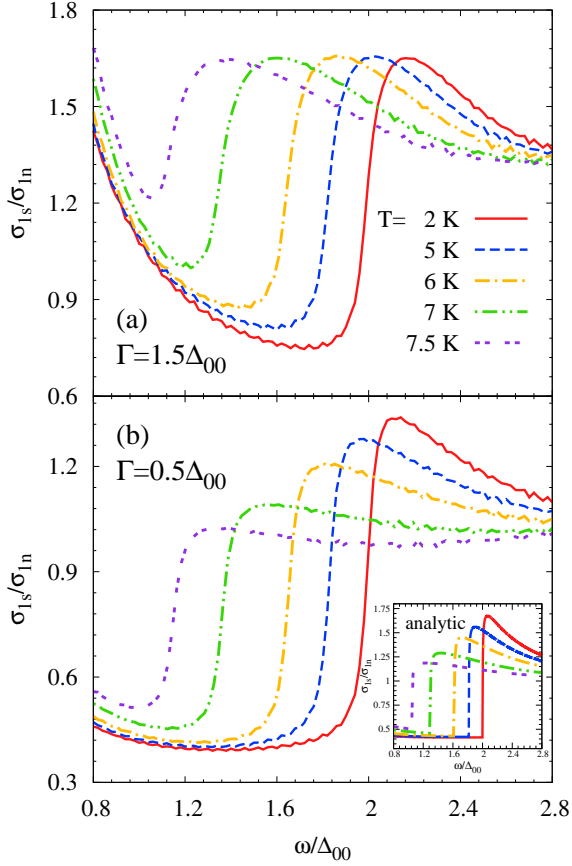}}
 \caption{(Color online) $\omega$ vs $\sigma_{1s}(\omega)/\sigma_{1n}(\omega)$
   at different temperatures by numerically calculating Eq.~(\ref{r1})
   for (a) $\Gamma_p=1.5\Delta_{00}$ and (b) $\Gamma_p=0.5\Delta_{00}$.
   In our calculation, $\Delta_0$ is calculated from Eq.~(\ref{gap0}).
   $\Delta_{00}$ denotes the order parameter at zero temperature. 
   Other parameters used in our calculation are listed in
   Table~\ref{Table}. The inset in (b) shows analytic solution from
   Eq.~(\ref{sigma1}).}
  
 \label{figyw4}
 \end{figure}
 
\section{Optical absorption in the dirty limit}
\label{FNC}

By full numerical calculation of Eq.~(\ref{r1}), the
frequency dependence of the optical absorption towards the dirty
limit are plotted in Fig.~\ref{figyw4}. As seen from
Fig.~\ref{figyw4}(a), both the upturn of the finite
$\sigma_{1s}(\omega)$ below $2\Delta_0$ and the crossover point at
$\omega=2\Delta_0(T)$ in $\sigma_{1s}(\omega)$ appear in the
dirty-limit regime ({\small $\xi<l$}), justifying our analysis in
Sec.~\ref{LROC} and in qualitative agreement with the experimental
findings \cite{NSL1,NSL2,NSL10,NSL3,NSL4,NSL0,NSL11,NSL5,NSL6,NSL8,NSL9,NSL12}. 
Thus, the GIKE provides an efficient approach to capture the
optical conductivity in the normal-skin-effect region.
To quantitatively fit the experimental data in the dirty limit,
the specific parameters of the density, effective mass and
momentum-relaxation rate are necessary, and this goes beyond the
scope of the present work.  

\section{Derivation of Eq.~(\ref{DS})}
\label{cc}

We derive Eq.~(\ref{DS}) in this part. Following the approach in our previous work in the clean limit \cite{GOBE5}, the solution of $\rho^{2\omega}_{\bf k}$ from Eq.~(\ref{G2}) in
the presence of the scattering is written as 
\begin{eqnarray}
\rho^{2\omega}_{{\bf k}2}&=&i\omega{A_{\bf k}}+S^c_{\bf k},\\
\rho^{2\omega}_{{\bf k}1}&=&-\xi_k{A_{\bf k}}-\frac{\xi_kS^c_{\bf k}}{i\omega}+\frac{\partial_t\rho_{\bf k}|^{2\omega,\tau_1}_{\rm
    sc}}{2i\omega},\\
\rho^{2\omega}_{{\bf k}3}&=&\Delta_0A_{\bf k}+B_{\bf k}+\frac{\Delta_0S^c_{\bf k}}{i\omega}+\frac{\partial_t\rho_{\bf k}|^{2\omega,\tau_3}_{\rm
    sc}}{2i\omega},
\end{eqnarray}
where $\partial_t\rho_{\bf k}|^{2\omega,\tau_i}_{\rm
    sc}$ denotes the $\tau_i$ component of the scattering term $\partial_t\rho_{\bf k}|^{2\omega}_{\rm sc}$;
  $A_{\bf k}$, $B_{\bf k}$ and $S^c_{\bf k}$ are given by
\begin{eqnarray}
A_{\bf k}&=&\frac{a_{\bf k}\!-\!\Delta_0[(\frac{e{\bf E}_0}{i\omega}\!-\!{\bf p}_s)\cdot\partial_{\bf k}\rho^{\omega}_{{\bf
      k}0}\!-\!({\bf p}_s\cdot{\partial_{\bf k}})^2\rho^0_{{\bf k}3}/4]}{2(\omega^2-E_k^2)},~~~~~~\\
B_{\bf k}&=&-\frac{(e{\bf E}_0\cdot{\partial_{\bf k}})\rho^{\omega}_{{\bf
      k}0}}{2i\omega},\\
S^c_{\bf k}&=&\frac{\xi_k\partial_t\rho_{\bf k}|^{2\omega,\tau_1}_{\rm
    sc}\!+\!i\omega\partial_t\rho_{\bf k}|^{2\omega,\tau_2}_{\rm sc}\!-\!\Delta_0\partial_t\rho_{\bf k}|^{2\omega,\tau_3}_{\rm
    sc}}{2(E_k^2\!-\!\omega^2)}.
\end{eqnarray}
The scattering term $\partial_t\rho_{\bf k}|^{2\omega}_{\rm sc}$ [Eq.~(\ref{nscat})] reads:
\begin{eqnarray}
&&\partial_t\rho_{\bf k}|^{2\omega}_{\rm sc}=-n_i\pi\sum_{{\bf k'}}|V_{\bf k-k'}|^2[Y^0_{\bf
  kk'}(2\omega)(\rho^{2\omega}_{{\bf k}0}-\rho^{2\omega}_{{\bf k}0})\nonumber\\
&&\mbox{}+Y^3_{\bf
  kk'}(2\omega)(\rho^{2\omega}_{{\bf k}3}-\rho^{2\omega}_{{\bf k}3})+Y^1_{\bf
  kk'}(2\omega)(\rho^{2\omega}_{{\bf k}1}+\rho^{2\omega}_{{\bf k}1})\nonumber\\
&&\mbox{}+Y^2_{\bf
  kk'}(2\omega)(\rho^{2\omega}_{{\bf k}2}+\rho^{2\omega}_{{\bf k}2})].\label{2sc}
\end{eqnarray}

At the weak scattering, by substituting the clean-limit solution of 
$\rho_{\bf k}^{2\omega}$ into the scattering terms as the first-order iteration, 
Eq.~(\ref{2sc}) becomes
\begin{eqnarray}
&&\partial_t\rho_{\bf k}|^{2\omega}_{\rm sc}=-n_i\pi\sum_{{\bf k'}}|V_{\bf k-k'}|^2[\Delta_0Y^3_{\bf
  kk'}(2\omega)(A_{\bf k}-A_{\bf k'})\nonumber\\
&&\mbox{}+i{\omega}Y^2_{\bf
  kk'}(2\omega)(A_{\bf k}+A_{\bf k'})-Y^1_{\bf
  kk'}(2\omega)(\xi_kA_{\bf k}+\xi_{k'}A_{\bf k'})\nonumber\\
&&\mbox{}+Y^3_{\bf
  kk'}(2\omega)(B_{\bf k}-B_{\bf k'})].
\end{eqnarray}
Then, $\rho^{2\omega}_{\bf k}$ is solved.

Consequently, with $Y^i_{\bf kk'}(2\omega)$ given by
Eqs.~(\ref{Y0})-(\ref{Y2}), substituting  $\rho^{2\omega}_{{\bf k}1}$ into Eq.~(\ref{gap}), 
one has
\begin{eqnarray}
&&\frac{\delta|\Delta|^{2\omega}}{g}={\sum_{\bf k}}'\xi_k^2C_k-\frac{in_i\pi\omega}{4}\sum_{\bf
  kk'}\frac{\xi_k^2\xi^2_{k'}}{E_kE_{k'}}|V_{\bf k_F-k'_F}|^2\nonumber\\
&&\mbox{}\times\Bigg[\frac{C_{k'}\delta(E_{k'}+2\omega-E_k)}{(E_k+\omega)(E_{k'}+\omega)}+\frac{C_{k}\delta(E_{k'}+2\omega-E_k)}{(E_k-\omega)(E_{k'}-\omega)}\nonumber\\
&&\mbox{}+\frac{C_{k'}\delta(E_{k}+2\omega-E_{k'})}{(E_k-\omega)(E_{k'}-\omega)}+\frac{C_{k}\delta(E_{k}+2\omega-E_{k'})}{(E_k+\omega)(E_{k'}+\omega)}\nonumber\\
&&\mbox{}+\frac{C_{k'}\delta(E_{k'}+2\omega+E_k)}{(E_k-\omega)(E_{k'}+\omega)}+\frac{C_{k}\delta(E_{k'}+2\omega+E_k)}{(E_k+\omega)(E_{k'}-\omega)}\nonumber\\
&&\mbox{}+\frac{C_{k'}\delta(2\omega-E_{k'}-E_k)}{(E_k+\omega)(E_{k'}-\omega)}+\frac{C_{k}\delta(2\omega-E_{k'}-E_k)}{(E_k-\omega)(E_{k'}+\omega)}\Big],~~~~~~
\end{eqnarray}
where $C_k$ is given by
\begin{equation}
C_k=\frac{\delta|\Delta|^{2\omega}g(E_k)}{E_k^2-\omega^2}+\frac{\Delta_0v_F^2\Big(\frac{e{\bf
  E}_0}{i\omega}-\frac{{\bf p}^{\omega}_s}{2}\Big)^2d(E_k)}{6(E_k^2-\omega^2)}.
\end{equation}
Here, 
we have taken care of
the particle-hole symmetry to remove terms with the
odd orders of $\xi_k$ and $\xi_{k'}$ in the summation of ${\bf k}$ and ${\bf k'}$;
we also take $\eta_{\bf k}$ in $\rho^{\omega}_{\bf k}$ [Eq.~(\ref{sor1})]
as its average value ${\bar \eta}_{\bf k}$ in the momentum space.
Then, after the mathematical integral, Eq.~(\ref{DS}) is obtained.

\section{Solution of Eqs.~(\ref{one})-(\ref{three})}
\label{dd}

In this part, we analytically solve Eqs.~(\ref{one})-(\ref{three}). 
Considering the weak scattering, we only keep zeroth and first orders of the scattering strength $\Gamma_0$ in the
following derivation.  Similar to the Elliot-Yafet relaxation mechanism in the
semiconductor spintronics \cite{spintronic},
following the L\"{o}wdin partition method \cite{dia}, through 
a unitary transformation {\small $(\delta\rho^s_{k+},\delta\rho^s_{k-},\delta\rho^s_{k3})^T=(1-S)(\delta\rho^q_{k+},\delta\rho^q_{k-},\delta\rho^q_{k3})^T$} with 
\begin{equation}
S=\frac{{\rm sgn}(\xi_k)\Delta_0\Gamma_0}{2iE_k^2}\left(\begin{array}{ccc}
0&0&0\\
0&0&0\\
1&-1&0
\end{array}\right),
\end{equation}
Eqs.~(\ref{one})-(\ref{three}) become
\begin{eqnarray}
&&\partial_t\delta\rho^s_{k+}+(2iE_k\delta\rho^s_{k+}+\gamma_k)\delta\rho^s_{k+}=2ia_k,\\
&&\partial_t\delta\rho^s_{k-}-(2iE_k\delta\rho^s_{k-}-\gamma_k)\delta\rho^s_{k-}=-2ia_k,~~\\
&&\partial_t\delta\rho^s_{k3}=-2\frac{\Delta_0{\rm sgn}(\xi_k)\Gamma_0}{E_k^2}a_k,
\end{eqnarray}
from which $\delta\rho_k^s$ can be directly solved:
\begin{eqnarray}
\delta\rho^s_{k\pm}&=&-c_{k\pm}\exp[{-(\pm2iE_k+\gamma_k)t}]\pm\int^t_0{2ia_k(t')}\nonumber\\
&&\mbox{}\times\exp[{-(\pm2iE_k+\gamma_k)\delta{t}}]dt',\\
\delta\rho^s_{k3}&=&-c_{k3}-\int^t_0\frac{2\Delta_0{\rm sgn}(\xi_k)\Gamma_0}{E_k^2}a_k(t')dt'.
\end{eqnarray}
Through the inverse transformations {\small $\delta\rho_{k}=U^{\dagger}_k\delta\rho^q_kU_k$} and
{\small
$(\delta\rho^q_{k+},\delta\rho^q_{k-},\delta\rho^q_{k3})^T=(1+S)(\delta\rho^s_{k+},\delta\rho^s_{k-},\delta\rho^s_{k3})^T$}, one has
\begin{eqnarray}
\delta\rho_{k1}&=&\frac{\xi_k}{E_k}\delta\rho_{k1}^s+\frac{\Delta_0}{E_k}\Big[{\delta\rho^s_{k3}-\frac{{\rm
    sgn}(\xi_k)\Delta_0\Gamma_0}{E_k^2}}\delta\rho^s_{k2}\Big],~~~~\label{sr1}\\
\delta\rho_{k2}&=&\delta\rho^s_{k2},\label{sr2}\\
\delta\rho_{k3}&=&\frac{\xi_k}{E_k}\Big[{\delta\rho^s_{k3}-\frac{{\rm sgn}(\xi_k)\Delta_0\Gamma_0}{E_k^2}}\delta\rho^s_{k2}\Big]-\frac{\Delta_0}{E_k}\delta\rho^s_{k1}.~~~~
\end{eqnarray}

Finally, substituting Eq.~(\ref{sr1}) into Eq.~(\ref{gap}),  by taking care of
the particle-hole symmetry to remove terms with the
odd order of $\xi_k$ in the summation of ${\bf k}$, one obtains
\begin{eqnarray}
&&\frac{\delta|\Delta|}{g}=\sum_{\bf
    k}\Big\{\frac{\Delta_0}{E_k}c_{k3}+\frac{\xi_k}{E_k}e^{-\gamma_kt}[c_{k1}\cos(2E_kt+\phi_k)\nonumber\\
&&\mbox{} -c_{k2}\sin(2E_kt-\phi_k)]-\left[\gamma_kg(E_k)\frac{\Delta_0^2}{E_k^2}\right]\int^t_0\delta|\Delta|(t')dt'\nonumber\\
&&\mbox{}+2g(E_k)\frac{\xi_k^2}{E_k}\int^t_0\delta|\Delta|(t')\sin(2E_k\delta{t}+\phi_k)e^{-\gamma_k\delta{t}}dt'\Big\},\nonumber\\
\end{eqnarray}
where the phase shift {\small $\phi_k=\arctan[{\rm
      sgn}(\xi_k)/\xi_k\Delta^2_0\Gamma_0/E_k^2]$} can be neglected at
the weak scattering. Then, Eq.~(\ref{damping}) is derived. 

\section{Derivation of Eq.~(\ref{FHD})}
\label{DFHD}

In this part, we derive Eq.~(\ref{FHD}). To consider the long-time 
dynamic behavior of the Higgs mode, by approximately taking
the starting point of time as $-\infty$ in Eq.~(\ref{damping}), one has
\begin{eqnarray}
  &&\frac{\delta|\Delta|(t)}{g}=A_i-\Gamma_H\int^t_{-\infty}\delta|\Delta|(t')dt' 
  +\sum_{\bf
    k}2E_kg(E_k)\frac{\xi_k^2}{E_k^2}\nonumber\\
  &&\mbox{}\times\int^t_{-\infty}\delta|\Delta|(t')\sin(2E_k\delta{t})e^{-\gamma_k\delta{t}}dt'\Big\},
\end{eqnarray}
with $A_i=\sum_{\bf k}\frac{\Delta_0}{E_k}c_{k3}$ and $\Gamma_H=\sum_{\bf
  k}\big[\gamma_kg(E_k)\frac{\Delta_0^2}{E_k^2}\big]$.
    
In the frequency space
{\small $\delta|\Delta|(t)=\int\frac{d\Omega}{2\pi}\delta|\Delta|_{\Omega}e^{-i\Omega{t}+0^+t}$},
the above equation becomes
\begin{equation}
 A_{i\Omega}=\Big[\frac{1}{g}-\frac{\Gamma_H}{i\Omega}-\sum_{\bf
    k}\frac{4g(E_k)\xi^2_k}{4E_k^2-(\Omega+i\gamma_k)^2}\Big]\delta|\Delta|_{\Omega}.
\end{equation}
By using Eq.~(\ref{gap0}) to replace $g$, one has
\begin{eqnarray}
A_{i\Omega}&=&\Big[\sum_{\bf
  k}\frac{(2\Delta_0)^2-(\Omega+i\gamma_k)^2}{4E_k^2-(\Omega+i\gamma_k)^2}+\frac{i\Gamma_H}{\Omega}\Big]\delta|\Delta|_{\Omega}\nonumber\\
&=&\Big[D\int{d\xi}\frac{(2\Delta_0)^2-(\Omega+i{\bar
      \gamma})^2}{4\xi^2+4\Delta^2_0-(\Omega+i{\bar \gamma})^2}+\frac{i\Gamma_H}{\Omega}\Big]\delta|\Delta|_{\Omega}\nonumber\\
&=&\Big[\frac{D\pi}{2}{\sqrt{(2\Delta_0)^2-(\Omega+i{\bar \gamma})^2}}+\frac{i\Gamma_H}{\Omega}\Big]\delta|\Delta|_{\Omega}.
  \end{eqnarray} 

Consequently, the temporal evolution of the Higgs mode is given by
\begin{widetext}
\begin{eqnarray}
\delta|\Delta|(t)&=&\int\frac{d\Omega}{\pi}\frac{A_{i\Omega}}{D\pi}\frac{e^{-i\Omega{t}+0^+t}}{{\sqrt{(2\Delta_0)^2\!-\!(\Omega\!+\!i{\bar
        \gamma})^2}}\!+\!2i{
      \Gamma}_H/(\Omega{D\pi})}=\int\frac{d\Omega}{\pi}\frac{A_{i\Omega}}{D\pi}\frac{e^{-i\Omega{t}+0^+t}[{\sqrt{(2\Delta_0)^2\!-\!(\Omega\!+\!i{\bar
          \gamma})^2}}\!-\!2i
      \Gamma_H/(\Omega{D}\pi)]}{[(2\Delta_0)^2\!-\!(\Omega\!+\!i{\bar \gamma})^2]\!+\![2
    \Gamma_H/(\Omega{D}\pi)]^2}  \nonumber\\
&=&\int\frac{d\Omega}{\pi}\frac{A_{i\Omega}}{D\pi}\frac{e^{-i\Omega{t}+0^+t}[{\sqrt{(2\Delta_0)^2\!-\!(\Omega\!+\!i{\bar
              \gamma})^2}}]}{(2\Delta_0)^2\!-\!(\Omega\!+\!i{\bar \gamma})^2\!+\![2
    \Gamma_H/(\Omega{D}\pi)]^2}\!-\!\Gamma_H\int\frac{d\Omega}{\pi}\frac{A_{i\Omega}}{D\pi}\frac{2i{e^{-i\Omega{t}+0^+t}}/(\Omega{D\pi})}{(2\Delta_0)^2\!-\!(\Omega\!+\!i{\bar
            \gamma})^2\!+\![2
    \Gamma_H/(\Omega{D}\pi)]^2}.~~~~~~~~
\end{eqnarray}
\end{widetext} 
Considering the weak scattering, the second term on the right-hand
side of the above equation can be neglected. By keeping the zeroth and
first orders of the scattering, one obtains
\begin{eqnarray}
  \delta|\Delta|(t)&\approx&\int\frac{d\Omega}{\pi}\frac{A_{i\Omega}}{D\pi}\frac{e^{-i\Omega{t}+0^+t}}{\sqrt{(2\Delta_0)^2\!-\!(\Omega\!+\!i{\bar
        \gamma})^2}}\nonumber\\
&=&e^{-{\bar \gamma}{t}}\int^{\infty+i{\bar \gamma}}_{-\infty+i{\bar
      \gamma}}\frac{d\Omega}{\pi}\frac{A_{i\Omega}}{D\pi}\frac{e^{-i\Omega{t}}}{\sqrt{(2\Delta_0)^2\!-\!\Omega^2}}\nonumber\\
  &\sim&e^{-{\bar \gamma}{t}}\int^{\infty+i{\bar \gamma}}_{-\infty+i{\bar
      \gamma}}\frac{d\Omega}{\pi}\frac{e^{-i\Omega{t}}}{\sqrt{(2\Delta_0)^2\!-\!\Omega^2}}.\label{CBR}
\end{eqnarray}
It is noted that for the integrand in Eq.~(\ref{CBR}),  in the complex
plane of $\Omega$, there exist two branching points at
$\Omega=\pm2\Delta_0$. Then, similar to the previous work\cite{OD3}, after the
standard construction of the closed contour, one obtains
\begin{equation}
  \delta|\Delta|(t){\sim}{\pi}e^{-{\bar
      \gamma}{t}}\frac{e^{2i\Delta_0t}+e^{-2i\Delta_0t}}{\sqrt{4\Delta_0t}}={\pi}e^{-{\bar
      \gamma}{t}}\frac{\cos(2\Delta_0t)}{\sqrt{\Delta_0t}}. 
\end{equation}

\section{Response of NG mode}
\label{ee}

As mentioned in the introduction, in our latest work for the clean limit \cite{GOBE5},
a finite second-order response of the NG mode, free from the influence of the Anderson-Higgs mechanism, is predicted as a consequence
of charge conservation. An experimental scheme for this response is
further proposed based on Josephson junction.  
In this part, for completeness,
we study the influence of the scattering on this response during and after the THz pulse.

\subsubsection{Excitation of NG mode in second-order response} 

During the pulse, substituting  $\rho^{2\omega}_{{\bf k}2}$ into Eq.~(\ref{phase}), 
the NG mode can be self-consistently derived:
\begin{eqnarray}
(\omega+i\gamma_0)\mu^{2\omega}_{\rm eff}&=&\frac{\omega+i\gamma_1+i\gamma_M}{3}\Big(\frac{e{\bf E}_0}{i\omega}-{\bf
    p}^{\omega}_s\Big)\cdot\frac{e{\bf E}_0}{i\omega{m}}g_{\omega}\nonumber\\
&&\mbox{}+\Big(\frac{e{\bf E}_0}{i\omega}-\frac{{\bf p}^{\omega}_s}{2}\Big)^2\frac{\omega+i\gamma_2}{6m}l_{\omega}.\label{NGSO}
\end{eqnarray}
Here, {\small
  $g_{\omega}={{\int}d\xi_kl(E_k)z(E_k)}/[{\int{d}\xi_kg(E_k)z(E_k)}]$}
and {\small
  $l_{\omega}={{\int}d\xi_km(E_k)z(E_k)}/[{\int{d}\xi_kg(E_k)z(E_k)}]$}
with {\small $z(E_k)=(E^2_k-\omega^2)^{-1}$} and
{\small $m(E_k)=(2\xi_k\partial_{\xi_k}+1)l(E_k)$}. The scattering contributions are given by 
\begin{eqnarray}
&&\gamma_M=\frac{3\int{d}\xi_k{\bar \eta_k}z(E_k)}{2{\int}d\xi_kl(E_k)z(E_k)},\\
&&\gamma_0=\frac{4\Gamma_0}{\int{d\xi_k}{g(E_k)z(E_k)}}\Big\{\int^{\infty}_{\Delta_0}dEo(E)o(E+2\omega)\nonumber\\
&&\mbox{}\times\omega(E+\omega)[g(E+2\omega)z(E+2\omega)-g(E)z(E)]\nonumber\\
&&\mbox{}-\int^{2\omega\!-\!\Delta_0}_{\Delta_0}dE{o(E)o(E-2\omega)}{\omega(\omega-E)g(2\omega-E)}\nonumber\\
&&\mbox{}\times{z(2\omega-E)\theta(\omega-\Delta_0)}\Big\},\label{gamma0}
\end{eqnarray}
and $\gamma_1$ and $\gamma_2$ are determined via replacing function $g(x)$ on the right-hand side of Eq.~(\ref{gamma0}) by $l(x)$
and $m(x)$, respectively. It is noted that in the absence of the scattering, Eq.~(\ref{NGSO}) exactly reduces to the
clean-limit one revealed in our previous work \cite{GOBE5}.

Consequently, similar to the investigation of the Higgs mode in Sec.~\ref{SORH}, 
the scattering also causes a phase-shift in the second-order response of the NG mode. 
Nevertheless, this phase shift is hard to detect, 
differing from the measurable optical response of
the Higgs mode in Sec.~\ref{SORH}.\\

 \begin{figure}[htb]
   {\includegraphics[width=7.8cm]{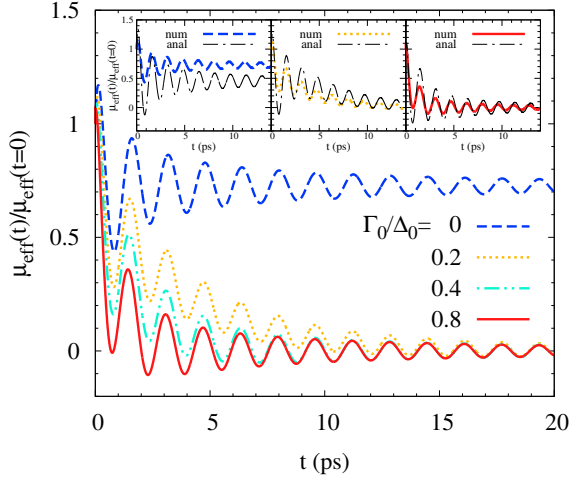}}
 \caption{(Color online)  Temporal evolution of NG mode $\mu_{\rm eff}$ after the optical pulse at different scattering
strengths.  The inset shows the
comparison between the analytic solution from Eq.~(\ref{damping}) and full numerical results from
Eq.~(\ref{SFDKE}). In the
calculation, $\delta\rho_{\bf k}(t=0)=\rho^{\omega}_{\bf k}+\rho^{2\omega}_{\bf k}$ with $\omega=\Delta_0$ and $T=~1$K. 
Other parameters used in our calculation are listed in Table~\ref{Table}.}
  
 \label{figyw5}
 \end{figure}

\subsubsection{Damping of NG mode}

After the pulse, by numerically solving our simplified model in Sec.~\ref{Simmodel}
[Eq.~(\ref{SFDKE}) with Eqs.~(\ref{gap})
and~(\ref{phase})], the temporal evolution of the optically excited NG mode is plotted in Fig.~\ref{figyw5} at different
scattering rates. As seen from the figure, the NG mode $\mu_{\rm eff}(t)=\partial_t\theta(t)$, i.e., the phase fluctuation, 
after the optical excitation
exhibits an oscillatory decay
behavior. The oscillating frequency is around $2\Delta_0$, and the
damping shows a monotonic enhancement with the increase of the
scattering rate.  Particularly, by further comparing
Figs.~\ref{figyw3} and~\ref{figyw5}, it is interesting to find that
the damping of the phase fluctuation (NG mode) is much faster than
that of the amplitude fluctuation (Higgs mode) of the order parameter.

Substituting the analytic solution 
of $\delta\rho_{k2}$ [Eq.~(\ref{sr2})] into Eq.~(\ref{phase}),  by taking care of
the particle-hole symmetry to remove terms with the
odd order of $\xi_k$ in the summation of ${\bf k}$, the analytic solution of $\mu_{\rm eff}(t)$ is derived:
\begin{eqnarray}
&&\int^t_0\mu_{\rm eff}(t')\sum_{\bf
  k}\Delta_0g(E_k)\cos(2E_k\delta{t})e^{-\gamma_k\delta{t}}dt'\nonumber\\
&&\mbox{}=-\sum_{\bf k}a_c\sin(2E_kt+\theta_c)e^{-\gamma_kt}.\label{damu}
\end{eqnarray}
As seen from Eq.~(\ref{damu}), 
terms on both the left- and right-hand sides show the oscillatory
decay with the time evolution, and hence, directly lead to the oscillating damping of $\mu_{\rm eff}(t)$ with the relaxation rate
proportional to $\Gamma_0$.
Comparisons between the solution from Eq.~(\ref{damu}) and the full numerical results from Eq.~(\ref{SFDKE}) are
plotted in the insets of Fig.~\ref{figyw5}, 
and the results from the two sets of calculations agree 
with each other again. 
    
\end{appendix}


\begin{thebibliography}{0}
\bibitem{vertex}  It is well established in normal metals that the vertex correction from the scatttering is inevitable
  to calculate the optical conductivity within the Kubo formalism, since without the vertex correction, one can not exactly
  derive the momentum relaxation time $\tau_p$.
\bibitem{GEG} In the quasiclassical approximation, Eilenberger derived a transport-like equation\cite{Eilen} from the gauge-invariant Gorkov's equation. 
  However, the gauge invariance is lost during his derivation.  Kita\cite{Kita} fixed this problem by introducing the Wilson line and constructed
  a gauge-invariant Eilenberger equation, in which complete
  electromagnetic effects are kept. Actually, all the
  electromagnetic terms emerged in the gauge-invariant Eilenberger
  equation also appear in our kinetic theory. Nevertheless, in Ref.~\onlinecite{CO}, the author used the original Eilenberger
  equation with incomplete electromagentic effect rather than the fixed gauge-invariant one.
\end{thebibliography}
\end{document}